\documentclass[]{aa}
\usepackage{natbib}
\usepackage{graphicx}
\usepackage{color}
\usepackage{amsmath}
\usepackage{url}

\bibpunct{(}{)}{;}{a}{}{,} 

\newcommand\bb[1] {   \mbox{\boldmath{$#1$}}  }

\definecolor{brown}{rgb}{0.42,0.24,0.07}
\definecolor{darkgreen}{rgb}{0.0,0.6,0.00}

\begin{document}


\title{Meridional circulation in turbulent protoplanetary disks}
\author{S\'ebastien Fromang\inst{1,2}, Wladimir Lyra\inst{3}, and
  Frederic Masset\inst{4,1,2}} 

\offprints{S.Fromang}

\institute{CEA, Irfu, SAp, Centre de Saclay, F-91191 Gif-sur-Yvette,
  France \and UMR AIM, CEA-CNRS-Univ. Paris VII, Centre de Saclay,
  F-91191 Gif-sur-Yvette, France. \and Department of Astrophysics,
  American Museum of Natural History, 79th Street at Central Park
  West, New York, NY 10024, USA \and Instituto de Ciencias F\'\i
  sicas, Universidad Nacional Aut\'onoma de M\'exico (UNAM),
  Apdo. Postal 48-3, 62251-Cuernavaca, Mor., Mexico \\ \email{sebastien.fromang@cea.fr}}

\date{Accepted; Received; in original form;}

\label{firstpage}

\abstract
{Based on the viscous disk theory, a number of recent studies have
suggested there is large scale meridional circulation in
protoplanetary disks. Such a flow could account for the presence of
crystalline silicates, including calcium-- and aluminum--rich
inclusions (CAIs), at large distances from the sun.} 
{This paper aims at examining whether such large--scale flows exist in
turbulent protoplanetary disks.}
{High--resolution global hydrodynamical and magnetohydrodynamical (MHD)
numerical simulations of turbulent protoplanetary disks were used
to infer the properties of the flow in such disks.}
{By performing hydrodynamic simulations using explicit viscosity, we
demonstrate that our numerical setup does not suffer from any
numerical artifact. The aforementioned meridional circulation is
easily recovered in viscous and laminar disks and is quickly
established. In MHD simulations, 
the magnetorotational instability drives turbulence in the
disks. Averaging out the turbulent fluctuations on a long timescale,
the results fail to show any large--scale meridional circulation. A
detailed analysis of the 
simulations show that this lack of meridional circulation is due to
the turbulent stress tensor having a vertical profile different from
the viscous stress tensor. A simple model is provided that
successfully accounts for the structure of the flow in the bulk
of the disk. In addition to those results, possible
deviations from standard vertically averaged $\alpha$ disk models are
suggested by the simulations and should be the focus of future work.}
{Global MHD numerical simulations of fully ionized and turbulent
  protoplanetary disks 
are not consistent with the existence of a large--scale meridional
flow. As a consequence, the presence of crystalline silicates at large
distance for the central star cannot be accounted for by that
process as suggested by recent models based on viscous disk theory.}

\authorrunning{S.Fromang et al.}
\titlerunning{Meridional flow}
\maketitle


\section{Introduction}
\label{introduction}

In the past few years, more and more evidence has accumulated that 
suggests there are crystalline solid particles in the outer
regions of protoplanetary disks. Spitzer observations in the $5$--$35$
$\mu$m spectral range have revealed crystalline signatures in a large
fraction of T--Tauri stars
\citep{bouwmanetal08,olofssonetal09,sargentetal09}. Comets, which are
believed to form far away from the Sun, are also known to show high
crystallinity values
\citep{crovisietetal97,woodenetal99,woodenetal07}. This has recently
been 
independently confirmed by the returned samples of the Stardust mission 
\citep{brownleeetal06,zolenskyetal06}. Meteoritic records also
indicate that calcium-- and aluminum--rich inclusions (CAIs) are a
common component of chondrites collected from parent bodies thought to
originate in the main asteroid belt. CAIs, as well as the crystalline
silicates observed by the Spitzer telescope, are believed to have
formed in the inner parts of protoplanetary disks ($R \leq 1$ AU)
where temperatures are in excess of $\sim$$1000$ K as required for their
formation from amorphous silicates. This general trend of finding
crystalline silicates at large distances from the central star
requires a mechanism able to transport them from the disk's inner parts
to their outer regions.

Several processes have been suggested to account for such 
transport. \citet{shuetal96,shuetal01} have invoked the action of
powerful winds from the young stars, the so--called X--wind
model \citep[see also][]{hu10}. Many authors have, on the
other hand, relied on processes operating within the bulk of the disk
itself. Indeed, the flow in accretion disks is believed to be 
turbulent in order for angular momentum to be efficiently transported
outward. The turbulent nature of the flow results in an efficient
diffusion of small solids in the disk, potentially transporting some
of them to large heliocentric distances. Using the standard 1D
(i.e. vertically integrated) $\alpha$
disk model \citep{shakura&sunyaev73,LP74} as a theoretical basis to
model the effect of the turbulence,
\citet{gail01} and \citet{bockeleemorvanetal02} have quantified the 
resulting concentration of crystalline silicates in the outer parts of
protoplanetary disks. The
conclusion is that turbulence alone appears unable to transport enough
solid material out to large distances because the inward flow
associated with mass accretion onto the central object dominates over
a long time and reduces its effect. As a remedy, several papers have
invoked the existence of a large--scale meridional 
flow in protoplanetary disks. Such a meridional circulation naturally
arises out of viscous disk models that extend the standard $\alpha$
disk model to two dimensions, the radial position $R$ and distance to
the midplane Z
\citep{urpin84,siemiginowska88,kley&lin92,rozyczkaetal94,regev&gitelman02,takeuchi&lin02}. In these models, the  
radial velocity is positive in the disk's equatorial plane while
gas accretion proceeds through the disk surfaces. This outwardly
directed mass flux advectively transports solids sedimented in the disk
equatorial plane out to large distances and thus circumvents the
limiting effect of mass accretion mentioned above. Several models based on
this idea have been developed in the last few years 
\citep{takeuchi&lin02,keller&gail04,tscharnuter&gail07,ciesla07,ciesla09,hughes&armitage10}. They
have indeed been able to successfully reproduce the amount of
crystalline solids found in outer protoplanetary disks. 

However, all of these models rely on the $\alpha$ prescription, a
large--scale model for the turbulence. While the nature of the
turbulence in protoplanetary disks is still somewhat debated, it is
most likely magnetohydrodynamical (MHD) in nature and driven by the
magnetorotational instability
\citep[MRI,][]{balbus&hawley91,balbus&hawley98}. Thanks to the large
increase in computational resources, it is now possible to perform
global numerical simulations of protoplanetary disks
\citep{pap&nelson03a,fromang&nelson06,lyraetal08} and to apply such
simulations to a variety of issues related to planet formation,
including 
dust dynamics \citep{fromang&nelson05,lyraetal08,fromang&nelson09},
planetesimals evolution \citep{nelson05,nelson&gressel10}, dead--zone
properties \citep{nataliaetal10}, and planet--disk interaction
\citep{nelson&pap03b,papetal04,nelson&pap04b,wintersetal03}. Using such
simulations, the nature of the large--scale flow can be investigated
from first principles without having to rely on ad hoc 
modeling of the turbulence. It is the purpose of the present paper to
develop such dedicated numerical simulations in order to validate the
models of meridional circulation presented above that are used to explain
the presence of crystalline solids at large distances from the central
objects in protoplanetary disks.

The plan of the paper is as follows. In
section~\ref{large_scale_flow}, we describe the properties of 
meridional circulation as it can be derived from 2D viscous disk
theory. Section~\ref{mhd_simus} presents a set of numerical
simulations of protoplanetary disks and analyzes the properties of the
large--scale flow. In section~\ref{num_checks_sec}, an additional series
of purely hydrodynamical simulations are used to assess the influence
of the numerical setup and algorithm 
on meridional circulation. Finally, in section~\ref{discussion_sec},
the results are discussed using a simple model and their consequences
for crystalline silicates radial mixing are highlighted.

\section{Large scale flow in protoplanetary disks}
\label{large_scale_flow}

In this section, we seek to derive the equations that govern the
large--scale radial flow in protoplanetary disks. To do so, we
consider an axisymmetric disk and derive its global
structure. The large--scale 
gas flow is determined by the mechanism governing angular momentum
transport. Two cases will be considered. We first examine the
case of viscous disk models in
section~\ref{ang_mom_visc_sec} and adopt the standard $\alpha$ prescription for
the kinematic viscosity. The analysis we present largely
follows the equations derived, for example, by \citet{takeuchi&lin02}
to which the reader is referred 
for further details. Since the flow in protoplanetary disk is most
likely turbulent, we also write in section~\ref{ang_mom_turb} the
equations that govern angular momentum transport in such disks,
highlighting the similarities and differences with viscous disk
models.

\subsection{Definitions}
\label{definitions}

Unless otherwise stated, we consider in this paper a cylindrical
coordinate system $(R,\phi,Z)$ with unit vectors
$(\bb{e_R},\bb{e_{\phi}},\bb{e_{Z}})$. The
disk model is fully specified once the equation of state (EOS),
midplane density, and viscosity are chosen. For the former, we used a 
locally isothermal EOS: the sound speed is time independent and given
as a function of the cylindrical radius $R$ by the power law, 
\begin{equation}
c_s^2=c_0^2\left(\frac{R}{R_0}\right)^q \, ,
\label{sound_speed}
\end{equation}
where $c_0$ is the sound speed at the fidutial radius $R=R_0$. The disk midplane
density $\rho_{mid}$ is similarly given by
\begin{equation}
\rho_{mid}=\rho_0 \left(\frac{R}{R_0}\right)^p \, ,
\label{midplane_rho}
\end{equation}
in which $\rho_0$ stands for the gas density at $R=R_0$. 

\subsection{Density and angular velocity}
\label{disk_setup_sec}

The first step is to derive the spatial distribution of gas density and
angular velocity. They are given by the equations of force balance in
the radial and vertical direction: 
\begin{eqnarray}
R \Omega^2-\frac{GMR}{(R^2+Z^2)^{3/2}}-\frac{1}{\rho}\frac{\partial
  P}{\partial R} &=& 0 \, , \label{force_bal_r}  \\
-\frac{GMZ}{(R^2+Z^2)^{3/2}}-\frac{1}{\rho}\frac{\partial
  P}{\partial Z} &=& 0 \, . \label{force_bal_z}
\end{eqnarray}
The second equation gives the vertical profile for the density:
\begin{eqnarray}
\lefteqn{ \rho(R,Z) = \rho_0 \left(\frac{R}{R_0}\right)^p \exp \left( \frac{GM}{c_s^2} \left[
    \frac{1}{\sqrt{R^2+Z^2}}-\frac{1}{R} \right] \right)\, , } 
\label{rho_rz}
\end{eqnarray}
where Eq.~(\ref{midplane_rho}) has been used to write the disk
midplane density. Upon expanding Eq.~(\ref{rho_rz}) to the second
order in $Z/R$, one recovers the standard Gaussian vertical profile
\begin{eqnarray}
\lefteqn{ \rho(R,Z) = \rho_0 \left(\frac{R}{R_0}\right)^p \exp \left(
    - \frac{Z^2}{2 H^2} \right) \, , } 
\label{rho_rz_gauss}
\end{eqnarray}
where $H=c_s/\Omega_K$ is the disk scale height, which is the ratio of
the sound 
speed to the Keplerian angular velocity $\Omega_K=\sqrt{GM/R^3}$. In
this last definition, $G$ and $M$ stand for the gravitational constant
and the central stellar mass, respectively . With the definition of
$c_s$ given in section~\ref{definitions}, $H$ can be 
expressed as
\begin{equation}
H=H_0 \left(\frac{R}{R_0}\right)^{(q+3)/2} \, ,
\end{equation}
where $H_0=c_0/\sqrt{GM/R_0^3}$ is the disk scale height at
$R=R_0$. Finally, Eq.~(\ref{force_bal_r}), along with the expression for
the gas density, gives the spatial variations of $\Omega$:
\begin{eqnarray}
\Omega &=& \Omega_K \left[ (1+q) - \frac{qR}{\sqrt{R^2+Z^2}} +
  (p+q) \left( \frac{H}{R} \right)^2 \right]^{1/2} \nonumber \\
&=& \Omega_K \left[ 1
  +\frac{1}{2} \left(\frac{H}{R}\right)^2 \left( p+q+ \frac{q}{2}
    \frac{Z^2}{H^2} \right) \right] \, . 
\label{omega_rz}
\end{eqnarray}
where the last equality results from a second--order expansion in
$Z/R$.

These expressions for the density and the angular
velocity do not depend on the viscous or turbulent 
nature of the flow. In particular, they are expected to hold in
turbulent protoplanetary disks provided turbulent fluctuations
are properly averaged out on long timescales, as well as in viscous
accretion disks regardless of the form of the viscous stress
tensor. We show in section~\ref{flow_prop} that this is indeed
the case.

\subsection{Angular momentum conservation in viscous disks}
\label{ang_mom_visc_sec}

A widespread and convenient way of modeling the effect of disk
turbulence is to solve the hydrodynamic equations adding a nonzero
kinematic viscosity. Since the early $1970$'s, this has led to the
development of $\alpha$ disk models in which the kinematic viscosity
is assumed to be of the form \citep{shakura&sunyaev73,LP74}
\begin{equation}
\nu=\alpha c_s H \, .
\label{alpha_prescription}
\end{equation}
In this section, we examine the consequences of such a prescription on
the disk's large--scale radial flow. Angular momentum conservation writes
in this case as
\begin{eqnarray}
\lefteqn{\rho \left( v_R\frac{\partial}{\partial R} +
v_Z\frac{\partial}{\partial Z} \right)(R^2 \Omega)} \nonumber \\
& & \hspace{1.5cm} = \frac{\partial }{\partial R} \left( R^2 T_{R
    \phi}^{visc}\right) + \frac{\partial}{\partial Z} \left(
  R^2 T_{Z \phi}^{visc} \right) \, ,
\label{ang_mom_visc}
\end{eqnarray}
where the viscous stress components $T_{R \phi}^{visc}$ and
$T_{Z \phi}^{visc}$ are given by 
\begin{eqnarray}
T_{R \phi}^{visc} &=& \rho \nu R \frac{\partial
  \Omega}{\partial R} \label{rphi_visc} \\
T_{Z \phi}^{visc} &=& \rho \nu R \frac{\partial
  \Omega}{\partial Z} \label{zphi_visc} \, .
\end{eqnarray}
In addition, we have the orderings $v_Z$\,$\sim$\,$(H/R) v_R$ and $\partial
/\partial Z$\,$\sim$\,$(H/R)\partial /\partial R$. Thus, in 
Eq.~(\ref{ang_mom_visc}), the term $v_Z \partial/\partial Z$ is a
factor $(H/R)^2$ smaller than the term $v_R \partial/\partial R$ and
can be safely neglected in thin disks for which $H$$\ll$$R$
\citep{takeuchi&lin02}. This gives 
\begin{eqnarray}
\lefteqn{ R\rho v_R\frac{\partial R^2 \Omega}{\partial R} =
\frac{\partial}{\partial R} \left( R^3\rho \nu \frac{\partial
  \Omega}{\partial R}\right) + \frac{\partial}{\partial Z} \left( R^3\rho \nu \frac{\partial
  \Omega}{\partial Z}\right) \, .}
\label{ang_mom_visc_II}
\end{eqnarray}
At this point, an expansion to the second order in $(Z/R)$ is performed in
order to proceed further in the analysis. After some
manipulations and using Eq.~(\ref{alpha_prescription}) for
the kinematic viscosity, the radial velocity can be expressed as
\citep{takeuchi&lin02}
\begin{eqnarray}
\lefteqn{\frac{v_R}{c_0}=-\alpha \left( \frac{H_0}{R_0} \right) \left(
\frac{R}{R_0} \right)^{q+1/2}} \nonumber  \\
& & \hspace{2cm} \left[ 3p+2q+6 + \frac{5q+9}{2} \left(  \frac{Z}{H} \right)^2\right] \, .
\label{vr_viscus_th}
\end{eqnarray}

In the disk midplane, the radial velocity is thus positive whenever
$3p+2q+6<0$, which is readily 
obtained for standard disks parameters. For example, typical values
generally considered in numerical simulations are $q=-1$ and $p=-3/2$
\citep{fromang&nelson06, fromang&nelson09, lyraetal09, nataliaetal10}, 
and they correspond to outward midplane velocities. For these typical
parameters, the second term appearing in front of the $Z$
dependence is positive. This indicates that the flow direction
reverses in the disk's upper layers to become negative. Overall, this
flow pattern produces a meridional circulation in the disk in which
gas flows outward in the disk midplane and accretes through the disk's
surface layers.

Equations~(\ref{rho_rz_gauss}), (\ref{omega_rz}), and
(\ref{vr_viscus_th}) completely determine the flow structure in a
viscous protoplanetary disk. It is established on a few dynamical
timescales, as suggested by \citet{takeuchi&lin02} and demonstrated
using 2D numerical simulations by \citet{rozyczkaetal94}. This does
not mean, however, that such a disk is in steady state. Indeed, that
is only the case for particular combinations of the exponents $p$ and
$q$ for which the mass flux is independent of radius 
\citep{takeuchi&lin02}. In a more general situation,
\citet{rozyczkaetal94} and \citet{kley&lin92} demonstrated using 2D
numerical simulation that a meridional circulation
is quickly established, on the one hand, as a dynamical response to the
viscous stress. On the other hand, the disk structure evolve viscously
on much longer timescales. We confirm these results using our own
simulations of viscous disks in Sect.~\ref{num_checks_sec}.

It is important at this stage to stress that the amplitude of the
radial velocity predicted by Eq.(\ref{vr_viscus_th}) is extremely
low. Indeed, as indicated by Eq.~(\ref{vr_viscus_th}), the Mach
number of the radial velocity is about $\alpha H_0/R_0$. With
standard values such as $\alpha \sim 10^{-2}$ and $H_0/R_0 \sim 0.1$,
this gives $v_R/c_0 \sim 10^{-3}$. In contrast, years of numerical
simulations of MRI--induced MHD turbulence have taught us that the
amplitude of the turbulent velocity fluctuations are around 
$5$ to $10$\% of the sound speed, i.e. almost two orders of magnitude
larger. Detecting the signature of a meridional circulation in
turbulent disk simulations will thus require extremely long
simulations to properly average out the turbulent velocity
fluctuations.

\subsection{Angular momentum conservation in turbulent disks}
\label{ang_mom_turb}

In a turbulent disk, angular momentum conservation can be written
in a form similar to Eq.~(\ref{ang_mom_visc_II}),
\begin{equation}
\label{ang_mom_turb_eq}
R\rho v_R\frac{\partial R^2 \Omega}{\partial R} =
\frac{\partial}{\partial R} \left( R^2 T_{R \phi}^{turb} \right) +
\frac{\partial}{\partial Z} \left( R^2 T_{Z \phi}^{turb} \right) \, ,
\end{equation}
but where $T_{R \phi}^{turb}$ and $T_{Z \phi}^{turb}$ now stand for the
turbulent stress tensors and are respectively given in terms of the
velocity and magnetic field turbulent fluctuations by the following
expressions \citep{balbus&pap99}:
\begin{eqnarray}
T_{R \phi} &=& <B_R B_{\Phi}-\rho v_R \delta v_{\phi}> \label{rphi_turb} \\
T_{Z \phi} &=& <B_Z B_{\Phi}-\rho v_Z \delta v_{\phi}> \, , \label{zphi_turb}
\end{eqnarray}
where $\delta v_{\phi}$ stands for the fluctuations of the azimuthal
component of the velocity. While analytical calculations have shown in
section~\ref{large_scale_flow} 
that the spatial variations in the viscous stresses $T_{R
  \phi}^{visc}$ and $T_{Z \phi}^{visc}$ lead to
large--scale meridional circulation in protoplanetary disks, no such
derivation can be performed for turbulent disks since their radial and
vertical profiles are unknown and highly fluctuating in time. One 
therefore has to rely on numerical simulations instead. In the following
sections, we present the results of such simulations and analyze the
effects of the turbulent stresses on the large--scale flow in turbulent
protoplanetary disks.

\section{Numerical simulations}
\label{mhd_simus}

\subsection{Setup}
\label{setup_mhd}

Three models of turbulent protoplanetary disks are presented in this
paper. The setup is very similar to those of
\citet{fromang&nelson06,fromang&nelson09}. The MHD equations are
solved with the code GLOBAL \citep{hawley&stone95} using spherical
coordinates $(r,\theta, \phi)$ with unit vectors
$(\bb{e_r},\bb{e_{\theta}},\bb{e_{\phi}})$.  The resolution is 
set to $(N_r,N_{\theta},N_{\phi})=(512,256,256)$. At time $t=0$, the gas
density and the angular velocity are initialized such that the disk
is in hydrostatic equilibrium. This is done using
Eq.~(\ref{rho_rz}) for the gas density and Eq.~(\ref{omega_rz}) for
the angular velocity. The radial and meridional velocities are both
set to zero. Units are such that $GM=1$, $R_0=1$, $\rho_0=1$, and
$c_0=0.1$. For all models, we used $q=-1$. This value gives a
steeper temperature radial profile than suggested by most realistic
disk models. It was largely chosen for computational reasons since it
gives a constant opening angle for the disk. Taken 
together, these 
relations mean that $H/R=H_0/R_0=0.1$ everywhere in the disk. The
computational grid covers the radial extent $r \in [1,10]$, the
azimuthal range $\phi \in [0,\pi/2]$, and five scale heights on both
sides of the equatorial plane: $\theta \in [\pi/2-0.5,\pi/2+0.5]$. The
grid resolution thus corresponds to more than $25$ cells per
scale height in the meridional direction. Finally, three values were
considered for the  
last parameter: $p=-3/2$, $-2$, and $-5/2$. They respectively give
a physically reasonable power law radial profile for the surface
density with exponents $-0.5$, $-1$, and $-1.5$. In the rest of
this paper, time is measured in orbital periods at the grid inner
edge. Using this set of parameters, each simulation can be recast into
physical units as explained in \citet{fromang&nelson09}.

To trigger the MRI at the beginning of each run, the
disk is initially threaded by a weak purely toroidal magnetic field
whose strength is such that the ratio $\beta$ 
between the thermal and magnetic pressure equals $25$
everywhere. Random velocity fluctuations are then added to each
component of the velocity, with an amplitude equal to $1\%$ of the
local sound speed. The boundary conditions are the same as used by
\citet{fromang&nelson06} and are periodic in azimuth and outflowing
in the meridional direction. The inner radial boundary uses the
"viscous outflow conditions", in which the radial velocity is set to a
constant value consistent with $\alpha$ disk theory, while all other
variables are outflowing. The outer radial boundary is handled through
a nonturbulent buffer zone (covering the range $R \in [9,10]$) where
motions and magnetic fields are damped.

With this setup, the disk quickly becomes fully turbulent because of
the MRI. The idea is then to average out the turbulent velocity
fluctuations in order to extract the mean flow structure of the
disk. This averaging is done both in space (using an azimuthal
average) and in time. However, as detailed in
section~\ref{ang_mom_visc_sec}, the  
expected amplitude of the meridional circulation is much smaller than
the amplitude of the turbulent velocity fluctuations. As a result,
the raw data need to be averaged over a long time interval to 
average these fluctuations out. We typically found that about
$400$ dumps evenly spaced over $200$ orbits at $R=1$ were needed to
extract a meaningful signal from the data. In the middle of the
computational domain, say at $R=5$, this corresponds to about $20$
orbits. This is much shorter than the viscous timescale (typically
a few hundred local orbital time) over which the 
disk evolves away from the initial state. Thus we expect that the disk
structure in the middle of the computational domain will remain close
to its initial state described by the power--law functions above. This
will facilitate the comparison with the models presented above.

An additional complication comes, however, from the 
violence of the MRI linear phase. Indeed, the transients
associated with it significantly affect the disk structure and drive 
it away from its initial state by the time the entire disk has become
fully turbulent. This prevents a detailed comparison with viscous
disk models. To solve that problem, the simulations were stopped after
$300$ orbits. At that time, the density and
velocity were reset to their initial values, keeping the magnetic
field to its current value. The models were then restarted for another
$300$ orbits. Time averaging was usually started about $100$ orbits
after this restarting procedure to avoid any spurious
transient that might be associated with it. As shown in the
next section, this procedure greatly helps to maintain the disk
close to its initial state and allows a clean comparison between
viscous and turbulent disk models.

\subsection{Fiducial run: p=-2}

We first present a detailed analysis of the model for which
$p=-2$, before using the other two cases to demonstrate the robustness
of the result. All the data presented in this section were obtained
after time averaging the raw simulation results between $t=400$ and
$t=600$ using roughly 400 snapshots evenly spaced in time by half an orbit.

\subsubsection{Flow properties}
\label{flow_prop}

\begin{figure*}
\begin{center}
\includegraphics[scale=0.4]{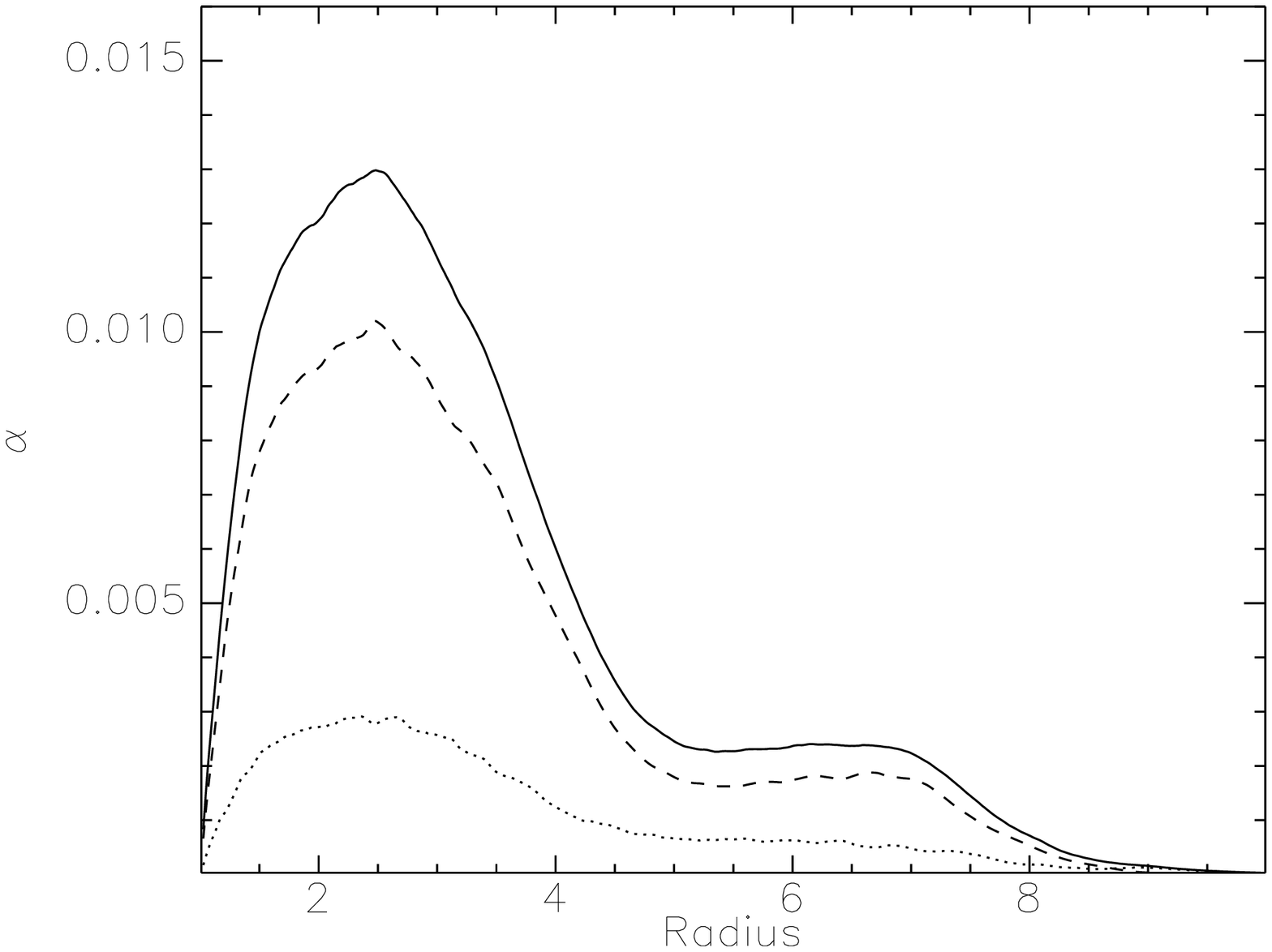}
\includegraphics[scale=0.4]{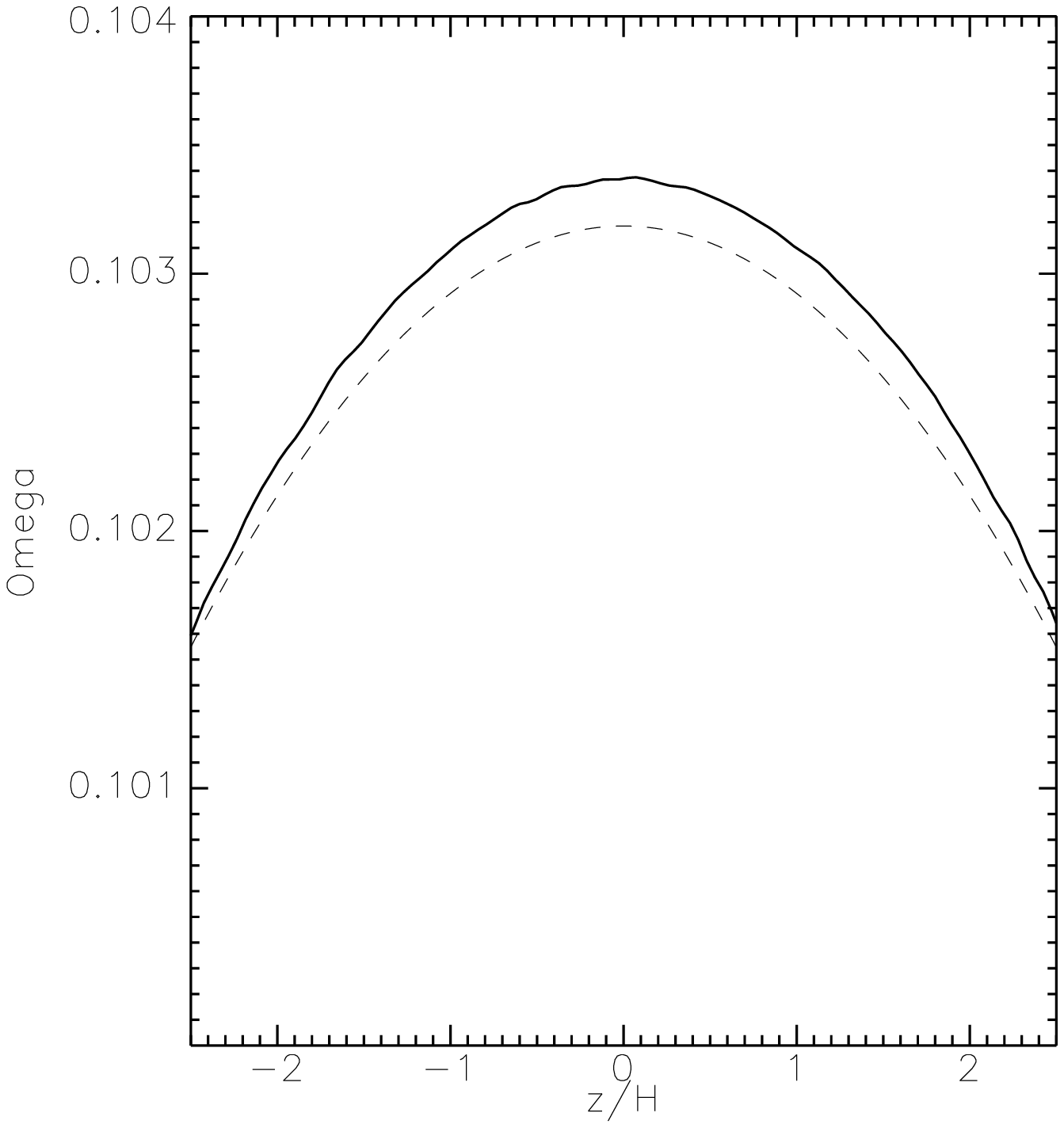}
\caption{Left panel: radial profile of $\alpha_{\rm Rey}$ ({\it dotted
   line}), $\alpha_{\rm Max}$ ({\it dashed line}), and $\alpha$ ({\it
   solid line}), as 
 defined according to Eq.~(\ref{alpha_eq}), for the case
 $p=-2$. Right panel: vertical profile of $\Omega$ at $R=4.5$ for the
 same model ({\it solid line}) compared with the prediction of
 Eq.~(\ref{omega_rz}) shown using the dashed line.}
\label{flow_prop_fig}
\end{center}
\end{figure*}

\begin{figure}
\begin{center}
\includegraphics[scale=0.5]{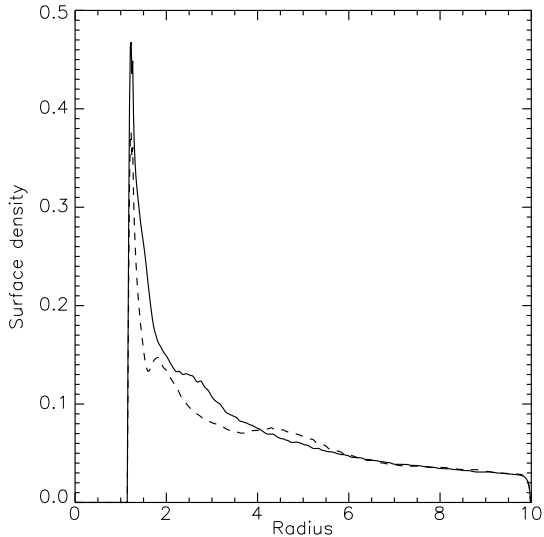}
\caption{Disk surface density for the case $p=-2$ at $t=400$ ({\it
    solid line}) and $t=600$ ({\it dashed line}).}
\label{surf_dens_fig}
\end{center}
\end{figure}

Before considering the possibility that a
meridional circulation develop in the disk, we first describe some
basic features of the 
flow that will be needed for the rest of the analysis. The most
important of them is provided by the parameter $\alpha$, which measures
the rate of angular momentum transport. As in
\citet{fromang&nelson06}, it is measured as a function of radius
according to 
\begin{equation}
\alpha(R)=\alpha_{\rm Rey}+\alpha_{\rm Max}=
\frac{\overline{ \delta v_R \delta v_{\phi}}-\overline{
   \frac{ B_R B_{\phi}}{4 \pi \rho}  }}{\overline{c_s^2}} \, ,
\label{alpha_eq}
\end{equation}
where $\alpha_{\rm Rey}$ and $\alpha_{\rm Max}$ correspond 
to the Reynolds and Maxwell stress contributions to $\alpha$, respectively.
The overbar symbols denote density-weighted azimuthal and vertical
averages. For example,
\begin{equation}
\overline{\delta v_R \delta v_{\phi}}=\frac{\int\int\rho \delta v_R
    \delta v_{\phi} d\phi dZ}{\int \int\rho d\phi dZ}=\frac{1}{2 \pi \Sigma}
\int\int\rho \delta v_R \delta v_{\phi} d\phi dZ \, ,
\end{equation}
where the last relation stands for a definition of the disk surface
density $\Sigma$.

The radial profile of $\alpha$ as defined above and time averaged
between $t=400$ and $t=600$ orbits is shown in
figure~\ref{flow_prop_fig} ({\it left panel}). In agreement with
published models of stratified protoplanetary disks
\citep{fromang&nelson06,fromang&nelson09,nataliaetal10,sorathiaetal10}, 
$\alpha$ is on the order of $0.005$. It shows fairly smooth variations with
$R$ that indicate a systematic decrease with radius. The reason for
such a radial decline is still unclear and beyond the scope of this
paper (see however a detailed discussion of that point in
section~\ref{comp_1D_alpha_sec}).

The right hand panel of figure~\ref{flow_prop_fig}
displays the vertical profile of $\Omega$ at radius $R=4.5$, averaged
azimuthally over the computational box and in time between $t=400$ and
$t=600$. For the purpose of comparison, the figure also displays the
prediction of Eq.~(\ref{omega_rz}). There is a small systematic
difference between the two curves because the
radial surface density profile is slightly flattened at $R=4.5$ over the
course of the simulation (see fig.~\ref{surf_dens_fig}), thereby
reducing pressure support and increasing $\Omega$. But aside from this
small difference, there is good agreement between the two curves. 
Finally, the quasi steady state structure of the disk is illustrated
by figure~\ref{surf_dens_fig}, which shows the radial profile of the
surface density at $t=400$ ({\it solid line}) and $t=600$ ({\it dashed
  line}). The two curves differ at most by $20$ to $30$\%, and the difference
is even much smaller in most parts of the disk. These
small variations in the disk surface density over the duration of the
simulations confirm that the disk structure remains close to
its initial state, as expected since the viscous timescale in the
middle of the computational domain is much longer than the duration of
the simulation. It can thus be accurately described using the power
laws we used in section~\ref{large_scale_flow}, which makes the
present simulations an excellent laboratory for studying whether
meridional circulation exists in turbulent protoplanetary disks. This
is the purpose of the following section.

\subsubsection{Meridional circulation}
\label{meridional_numerics}

\begin{figure*}
\begin{center}
\includegraphics[scale=0.335]{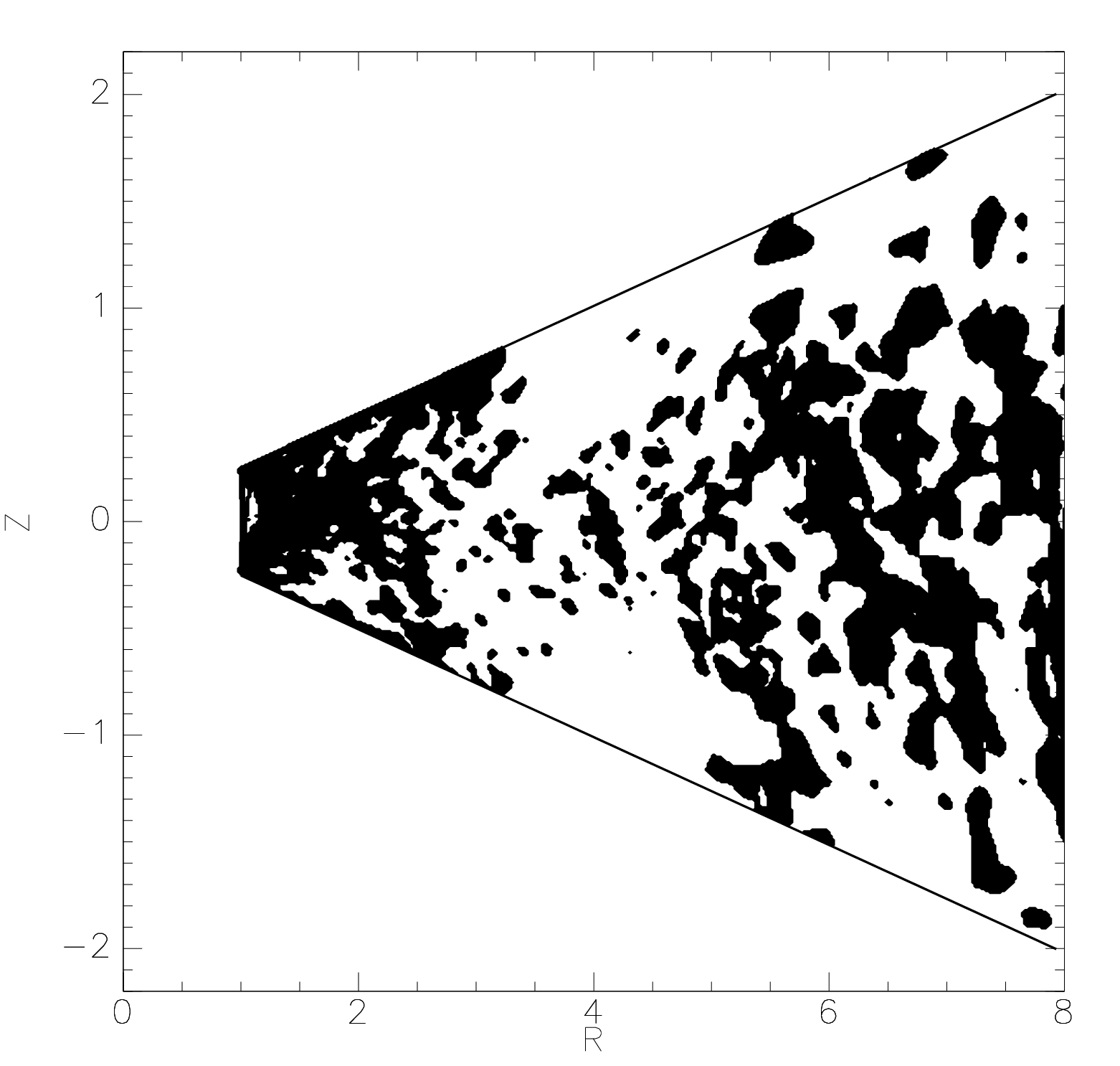}
\includegraphics[scale=0.45]{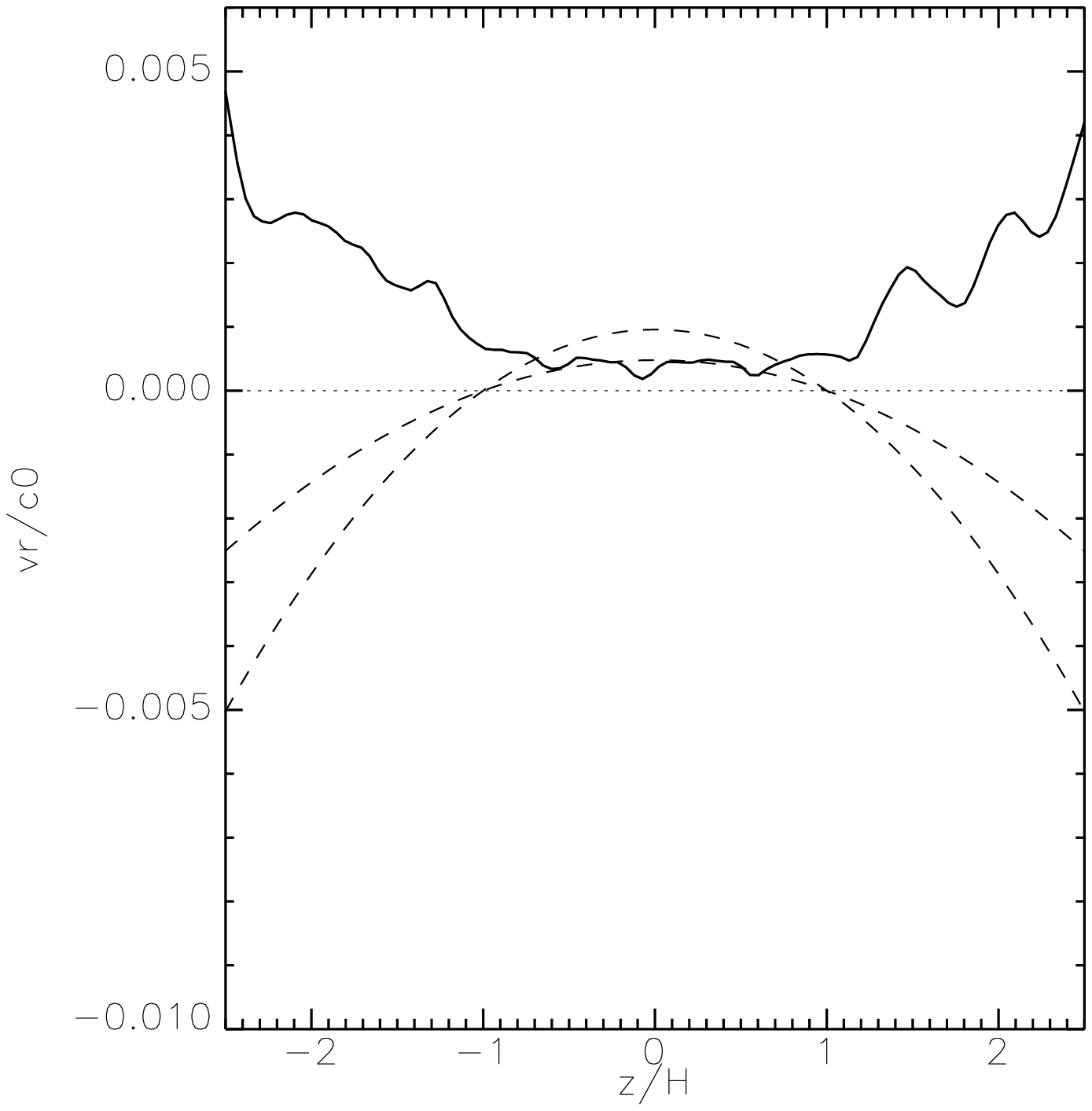}
\caption{Left panel: Time and azimuthally averaged radial velocity for
the model $p=-2$. Positive velocities are marked with white colors,
while black regions correspond to negative $v_R$. The raw simulation
data have been averaged in time between $t=400$ and $t=600$. Right
panel: The solid line shows the vertical profile of the radial
velocity averaged in time between $t=400$ and $t=600$, in the
azimuthal direction and in the radial direction between $R=3$ and
$R=6$. The dashed lines show the theoretical
prediction of Eq.~(\ref{vr_viscus_th}) for $\alpha=10^{-2}$ and $5 \times
10^{-3}$, respectively . The dotted line simply marks the zero point
as a reference.}
\label{case_p2_q1_fig}
\end{center}
\end{figure*}

\begin{figure*}
\begin{center}
\includegraphics[scale=0.45]{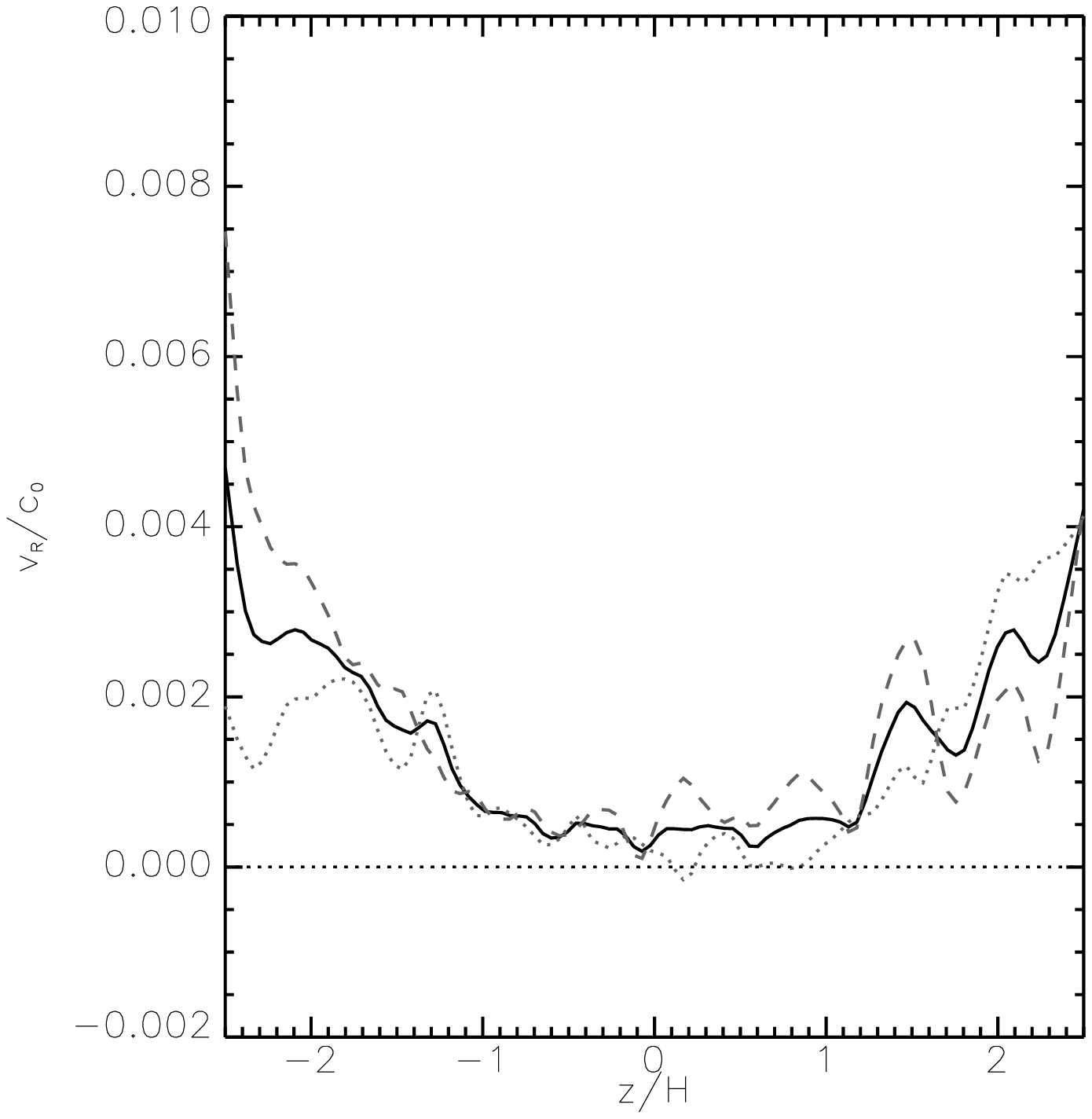}
\includegraphics[scale=0.45]{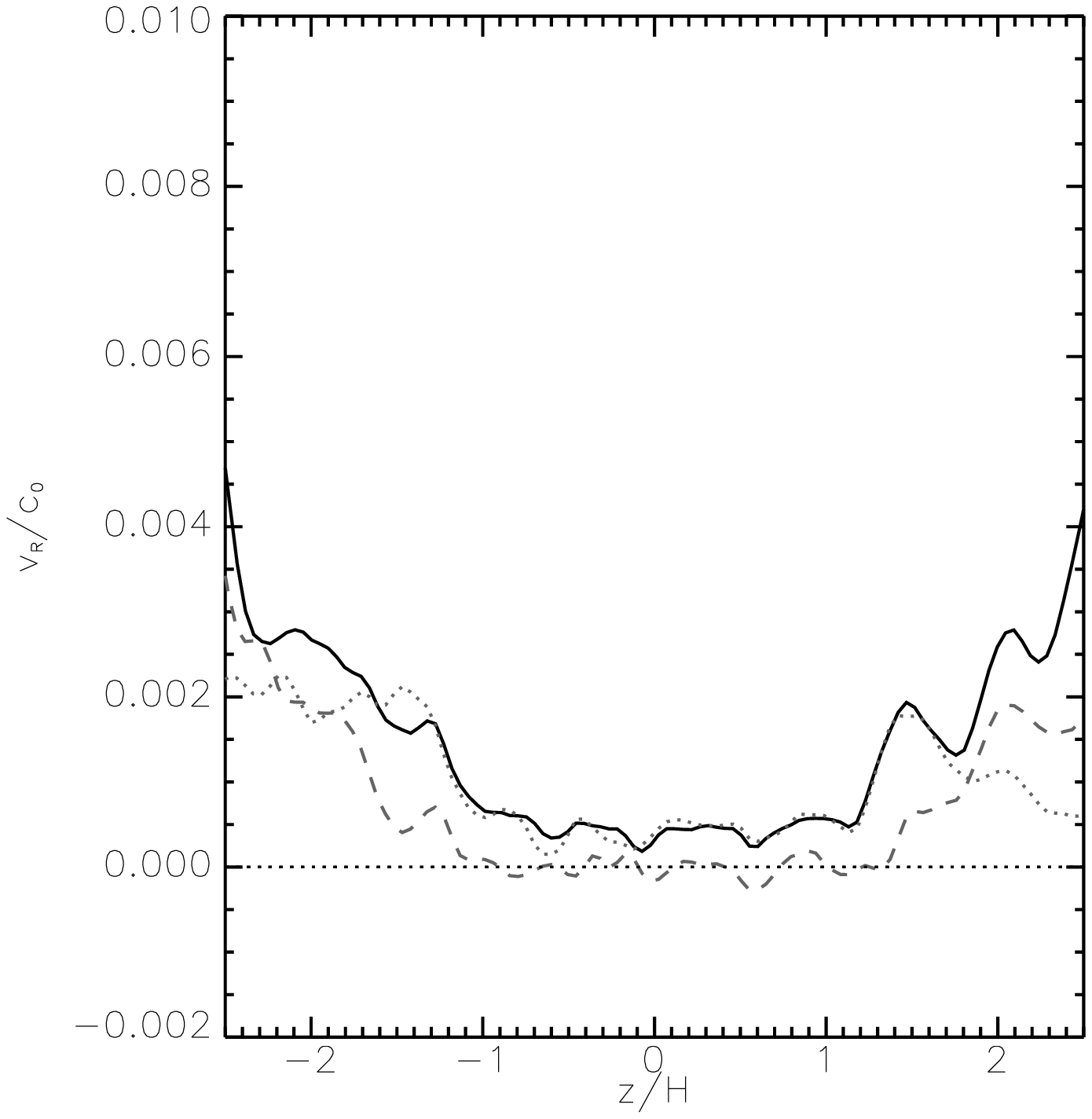}
\caption{Vertical profile of the radial velocity for the case
  $p=-2$. In both panels, the solid line is identical to that plotted
  on the right panel of figure~\ref{case_p2_q1_fig} (time average over
  $[400,600]$ and radial average over $[3,6]$). The left panel 
investigates the sensitivity of the result on the time average while
keeping the same radial range: the time average is taken over $[400,500]$
({\it dotted line}) and $[500,600]$ ({\it dashed line}). The right
panel shows data obtained with the same time average, $[400,600]$, but
different radial range, namely $R \in [2,4.5]$ ({\it dotted line}) and
$R \in [4.5,7]$ ({\it dashed line}).}
\label{case_p2_q1_figII}
\end{center}
\end{figure*}

The properties of the large--scale radial flow that develops during the
simulation are illustrated in figure~\ref{case_p2_q1_fig}. For the
results to be easier to interpret, only the region $|Z|<2.5H$ has been
considered in the analysis. This corresponds to the midplane region of
the disk where most of the mass concentrates. According to
Eq.~(\ref{vr_viscus_th}), this is also the region where meridional
circulation is expected to develop. Velocity fluctuations above that
height  
are so large that no converged mean flow could be extracted from the
simulations, even after averaging for $200$ orbits. In any case,
because of the exponential drop in density, little mass flux is
expected in that region. The sign
of the radial velocity in the $(R,Z)$ plane is displayed
in the left hand panel of figure~\ref{case_p2_q1_fig}. Despite the
very long averaging period, the patchy structure 
apparent in that figure indicates that the data remain very noisy even
after the long average we performed. As discussed above, this is due to the 
large--scale flow having an amplitude much smaller than the turbulent
velocity fluctuations. Nevertheless, it is
also evident from the figure that the numerical 
simulations do not show any obvious signature of a meridional
circulation. Such a feature would have been characterized by a large white
region around the disk midplane sandwiched between two black
regions. In fact, it appears that many black blobs are clustered
around the disk midplane, while the disk surface (i.e. the region
located at larger $|Z|$) appears whiter. This is confirmed by the
right hand panel of that figure, in 
which an additional radial averaging has been performed between $R=3$
and $R=6$. The resulting vertical profile of $v_R$ is compared with
the theoretical expectation of Eq.~(\ref{vr_viscus_th}) for two 
values of $\alpha$, namely $5 \times 10^{-3}$ and
$10^{-2}$. In the simulations, $v_R$ is
found to be low in the vicinity of the disk midplane 
and rises in the disk surface layers. It is surprising to see
positive velocities at all $Z$ since it indicates a mean bulk motion
of the disk. This will be largely addressed in
section~\ref{comp_1D_alpha_sec} where the connections between the
present simulations and vertically averaged $\alpha$--disks are
discussed. Here, we only comment that the large difference between
the simulations and the predictions of viscous disk theory further
supports the conclusion suggested above. There is no meridional
circulation in turbulent protoplanetary disks in which turbulence is
driven by the MRI. 

As mentioned above, the amplitude of the large--scale flow is
nevertheless much smaller than the typical turbulent velocity
fluctuations. This raises the question of the uncertainty on the
radial velocity vertical profile shown in the right hand panel of
figure~\ref{case_p2_q1_fig}. To investigate this point, we
have varied the intervals over which the time and radial averages are
performed. The result is shown in figure~\ref{case_p2_q1_figII}. In
both panels, the solid curve is identical to the one plotted in the
right hand 
panel of figure~\ref{case_p2_q1_fig}: the time average is performed
over the range $[400,600]$ and the radial average over the range
$[3,6]$. In the left hand panel, the dotted and dashed lines retain the
same radial range, $R \in [3,6]$, while the time interval 
is respectively varied over the range $[400,500]$ and
$[500,600]$. In the right hand panel, the 
time average is performed over the range $[400,600]$ while the radial
average is done over $[2,4.5]$ and $[4.5,7]$. As a result,
all of these additional curves 
are obtained by averaging the simulation results either over a smaller
radial extent or over a shorter time interval than used to compute the
solid curve. They are thus expected to
be less converged than the solid line. Nevertheless, the dotted and
dashed curves in both panels are qualitatively similar to the solid
line (i.e. no meridional circulation is found) and show only moderate
quantitative difference with the solid line, indicating that the
results shown in figure~\ref{case_p2_q1_fig} are converged fairly
well.

\subsubsection{Turbulent vs. viscous torques}
\label{torque_sec}

\begin{figure*}
\begin{center}
\includegraphics[scale=0.5]{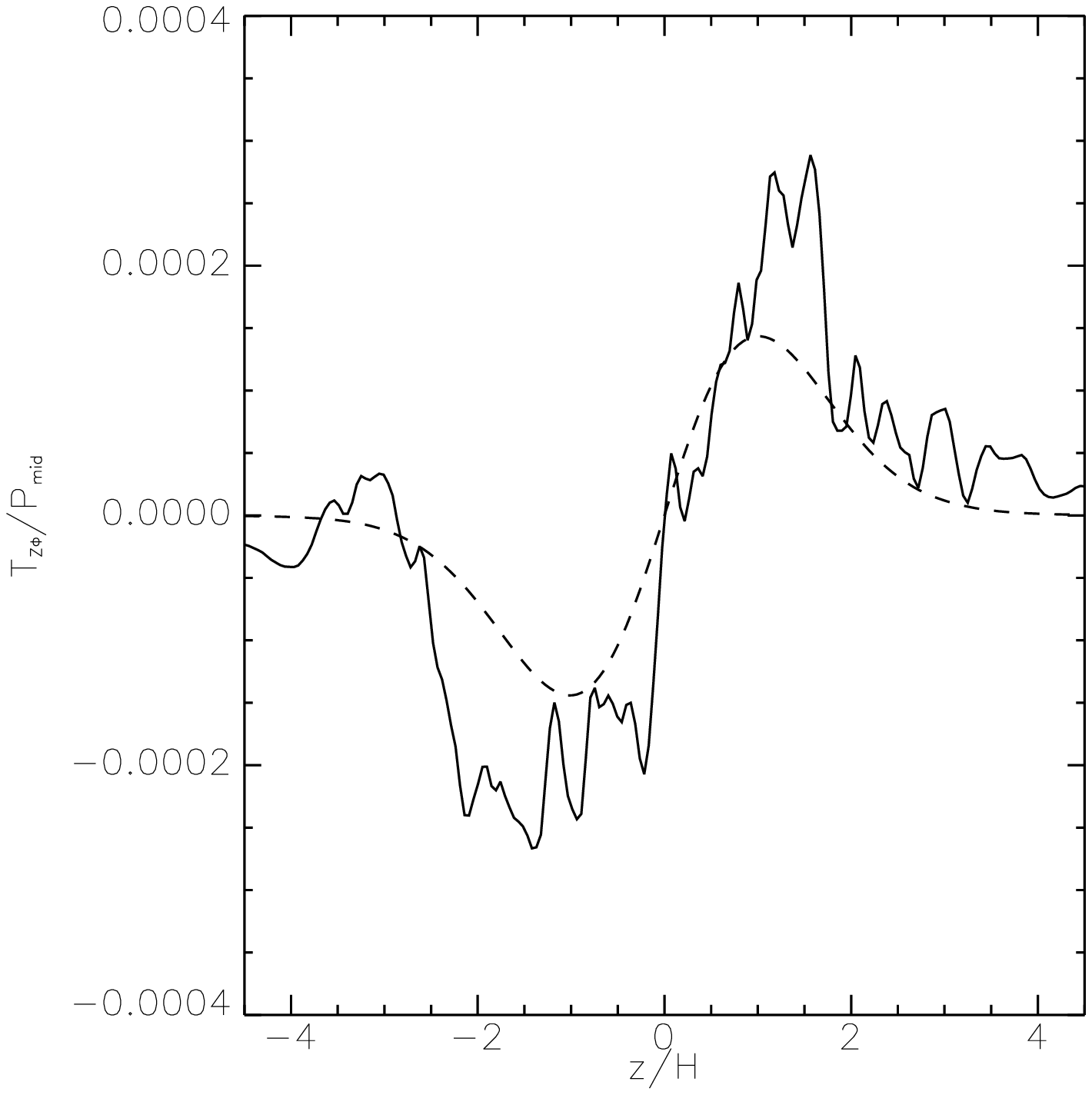}
\includegraphics[scale=0.5]{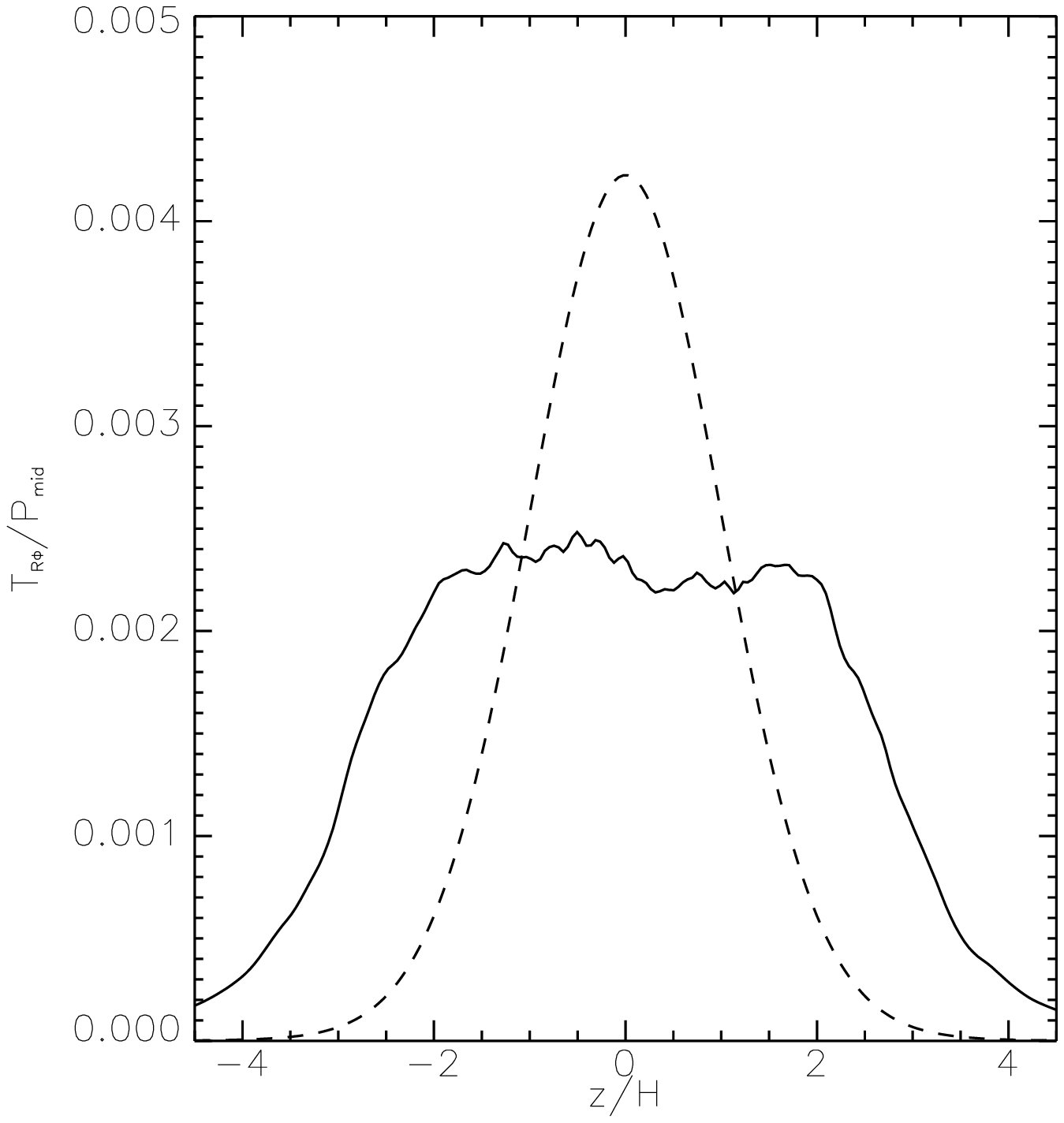}
\caption{Left panel: Vertical profile of $T_{Z\phi}^{turb}/P_{mid}$ ({\it
solid line}) and $T_{Z\phi}^{visc}/P_{mid}$ ({\it dashed line}), time
averaged between $t=400$ and $t=600$ and radially averaged between
$R=3$ and $R=6$. The two stresses are similar both in amplitude and in
their vertical profile. Right panel: Vertical profile of $T_{R\phi}^{turb}/P_{mid}$ ({\it 
solid line}) and $T_{R\phi}^{visc}/P_{mid}$ ({\it dashed line}), time
averaged between $t=400$ and $t=600$ and radially averaged between
$R=3$ and $R=6$. Although they are comparable in magnitude, the
viscous and turbulent stresses display largely different vertical profiles.} 
\label{1d_stress_p2_fig}
\end{center}
\end{figure*}

\begin{figure*}
\begin{center}
\includegraphics[scale=0.35]{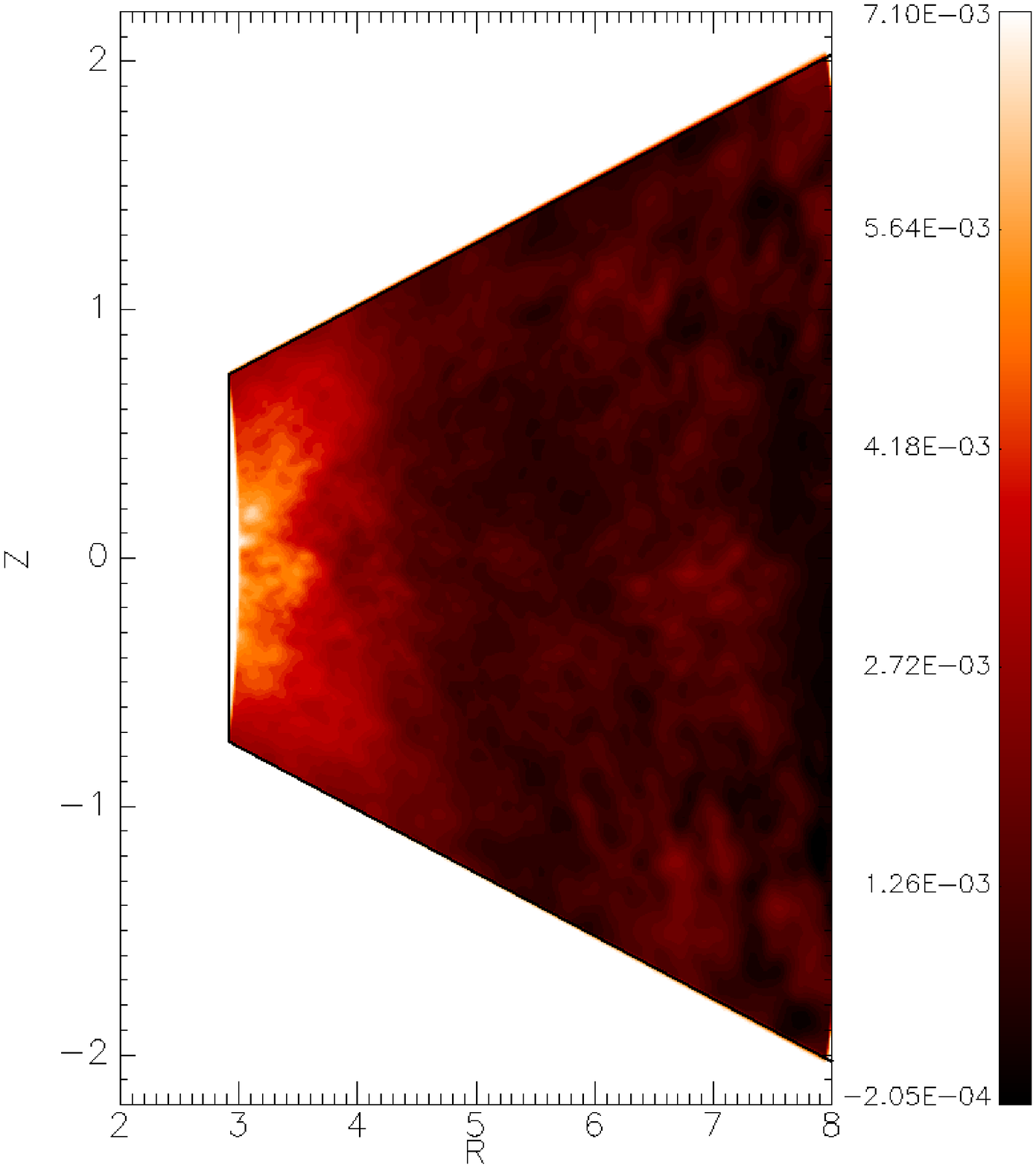}
\includegraphics[scale=0.35]{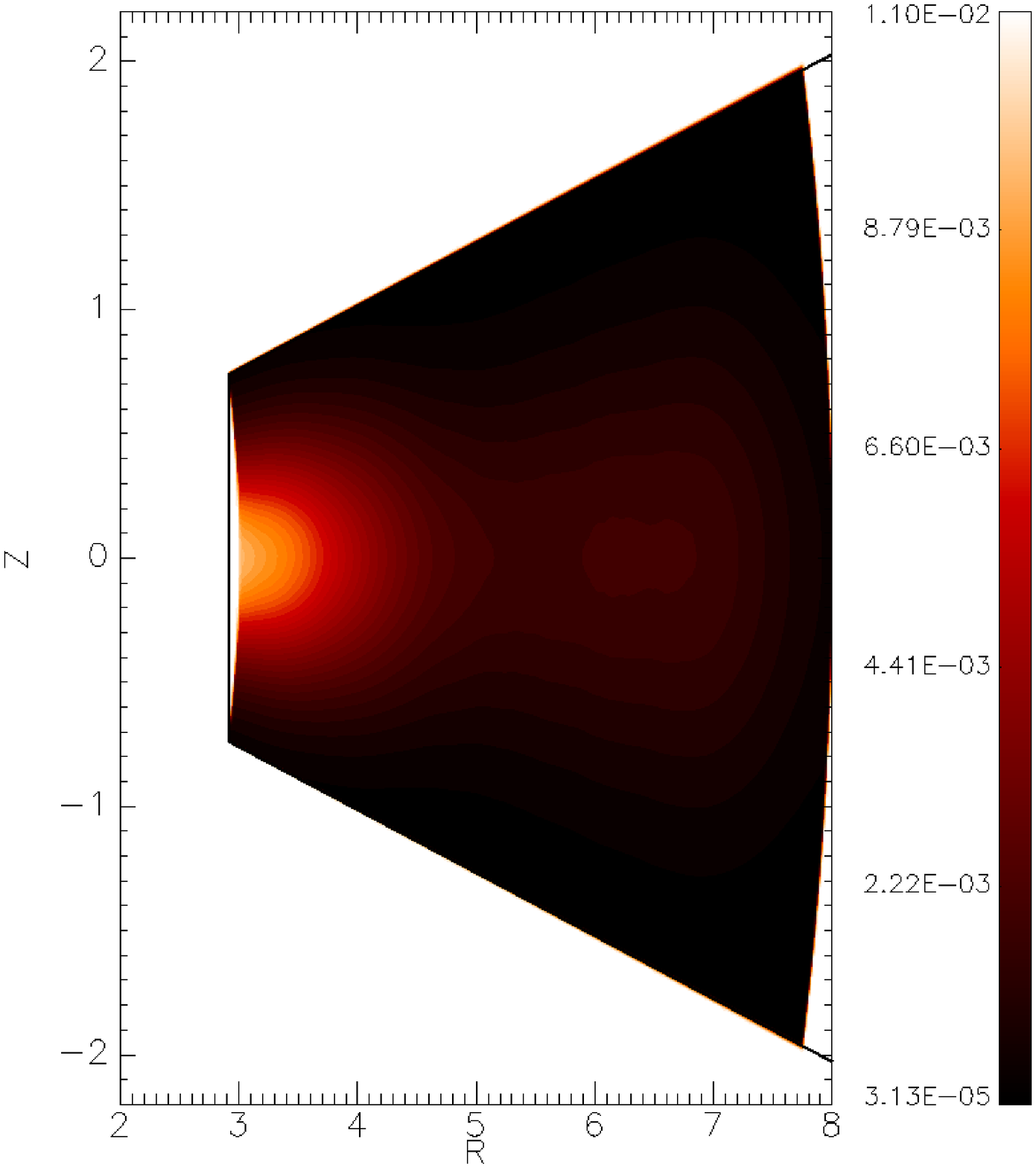}
\caption{Left panel: Spatial variation of $T_{R\phi}^{turb}/P_{mid}$ in the
disk's meridional plane, time--averaged between $t=400$ and
$t=600$. Right panel: Same as the left panel, but for the quantity
$T_{R\phi}^{visc}/P_{mid}$. A strong vertical stratification is apparent in
the viscous stress but appears to be much weaker in the turbulent stress.}
\label{2d_stress_p2_fig}
\end{center}
\end{figure*}

The results presented above demonstrate important differences between the
large--scale flow properties of viscous and turbulent disks. This is
due to differences between viscous and turbulent stresses. Indeed, as
outlined in section~\ref{ang_mom_turb}, stress 
tensors responsible for angular momentum transport have no reason to
be identical in viscous and in turbulent disks. We now compare them in
detail. 

We first consider the $Z\phi$ components of those stresses. Both
$T_{Z\phi}^{visc}$ and $T_{Z\phi}^{turb}$ are respectively given by
Eq.~(\ref{zphi_visc}), along with the $\alpha$--prescription for the
viscosity and Eq.~(\ref{zphi_turb}). The left hand panel of figure
\ref{1d_stress_p2_fig} compares the vertical 
profile of $T_{Z\phi}^{turb}/P_{mid}$ and
$T_{Z\phi}^{visc}/P_{mid}$. As for the right hand panel of 
figure~\ref{case_p2_q1_fig}, the simulation data have been averaged in
azimuth over the entire computational domain  and in radius between
$R=3$ and $R=6$. A further average over $11$ snapshots evenly spaced
between $t=300$ and $t=600$ was also performed\footnote{Unlike
the $R\phi$ component of the turbulent stress, the derivation of the
$Z\phi$ component of the turbulent stress had not been
anticipated before the simulations were performed. Their calculation
was done using the small number of dump files available after their
completion rather than during the simulations themselves as
for the $T_{R\phi}$ components. Although such a procedure
unfortunately results in larger fluctuations, it is not prohibitive
because stress tensors tend to converge faster than other statistical 
diagnostics like the mean radial velocity.}. The two stresses display
a comparable amplitude and a 
similar vertical profile, which is ultimately related to
the sign and the amplitude of the vertical derivative of
$\Omega$. 

We now turn our attention to the $R\phi$ components of the viscous and
turbulent stress tensors. These are given by
Eq.~(\ref{rphi_visc}), along with the $\alpha$--prescription for the
viscosity, and Eq.~(\ref{rphi_turb}), respectively. Snapshots of both
tensors in the disk's  
meridional plane, properly averaged in time and azimuth, are shown in
figure~\ref{2d_stress_p2_fig}. The left hand  
panel shows $T_{R\phi}^{turb}/P_{mid}$ while the right hand panel plots
$T_{R\phi}^{visc}/P_{mid}$ (as for the left panel of
figure~\ref{case_p2_q1_fig}, only the region within $2.5$H of the
equatorial plane has been plotted here). First, both the viscous and
turbulent $R\phi$ components of the stress
are much larger in amplitude than the $Z\phi$ components plotted in
the left hand panel. Second, their vertical profiless are markedly
different. While $T_{R\phi}^{turb}$ vertical variations are 
weak, those of $T_{R\phi}^{visc}$ are strong. This is because the
latter traces the large vertical gradient of the gas density, while the
former arises mainly because of magnetic forces. This
difference is made even more apparent in the right hand panel of
figure~\ref{1d_stress_p2_fig} in which a further radial average
between $R=3$ and $R=6$ has been performed. (The
range of the x--axis has been expanded up to $4.5H$ compared to
figure~\ref{2d_stress_p2_fig}.) In this figure, the solid line plots
the turbulent stress while  
the dashed line represents the viscous stress tensor. As suggested by
figure~\ref{2d_stress_p2_fig}, the viscous stress displays a vertical
profile reminiscent of the Gaussian density profile. The turbulent
stress, because it is dominated by the magnetic forces, shows a plateau for
$|Z|<2.5H$ before dropping to lower values higher in the
disk. Such a vertical profile for the turbulent stress tensor has
already been reported in numerous numerical simulations of MRI--driven
MHD turbulence, using either a local
\citep{miller&stone00,hiroseetal06,flaigetal10} or a global approach 
\citep{fromang&nelson06,nataliaetal10,sorathiaetal10}. As becomes 
clear in section~\ref{turb_disk_model}, it is precisely this peculiar
vertical structure in the turbulent stress that prevents meridional
circulation from developing.

\subsection{Dependence with the surface density power law index}

\begin{figure*}
\begin{center}
\includegraphics[scale=0.335]{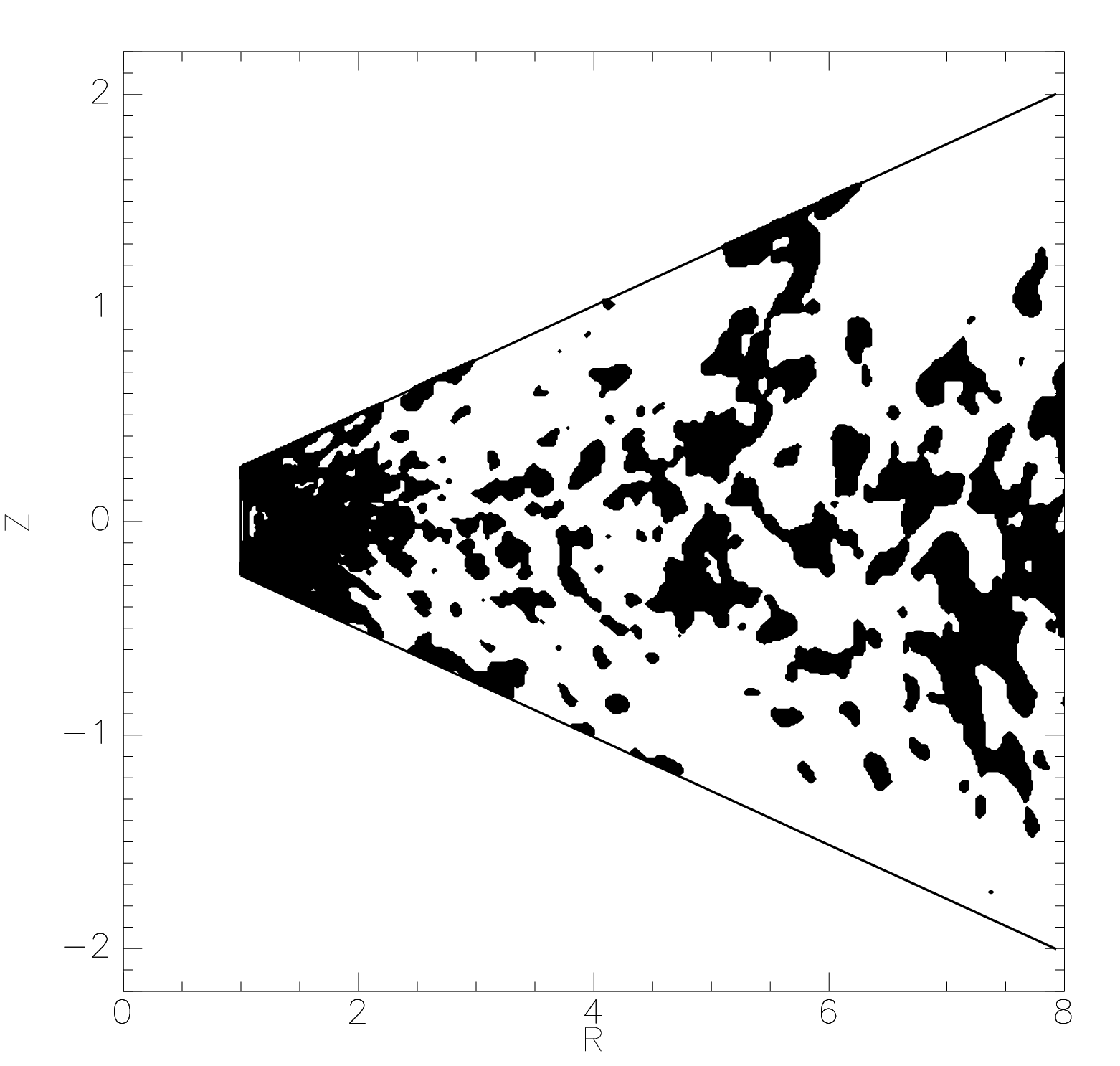}
\includegraphics[scale=0.45]{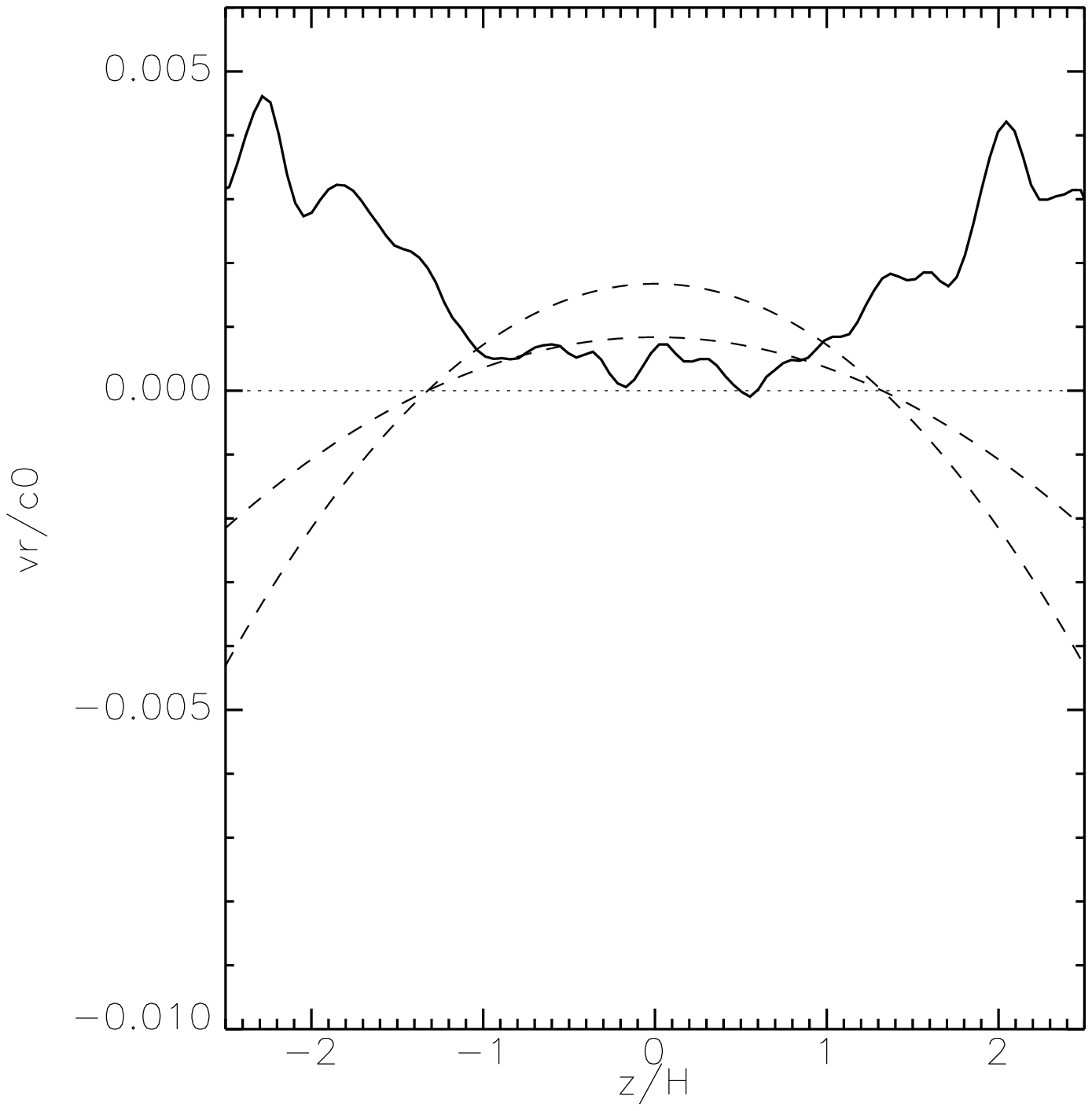}
\caption{Same as figure~\ref{case_p2_q1_fig} but for the case $p=-5/2$.}
\label{case_p2.5_q1_fig}
\end{center}
\end{figure*}

\begin{figure*}
\begin{center}
\includegraphics[scale=0.335]{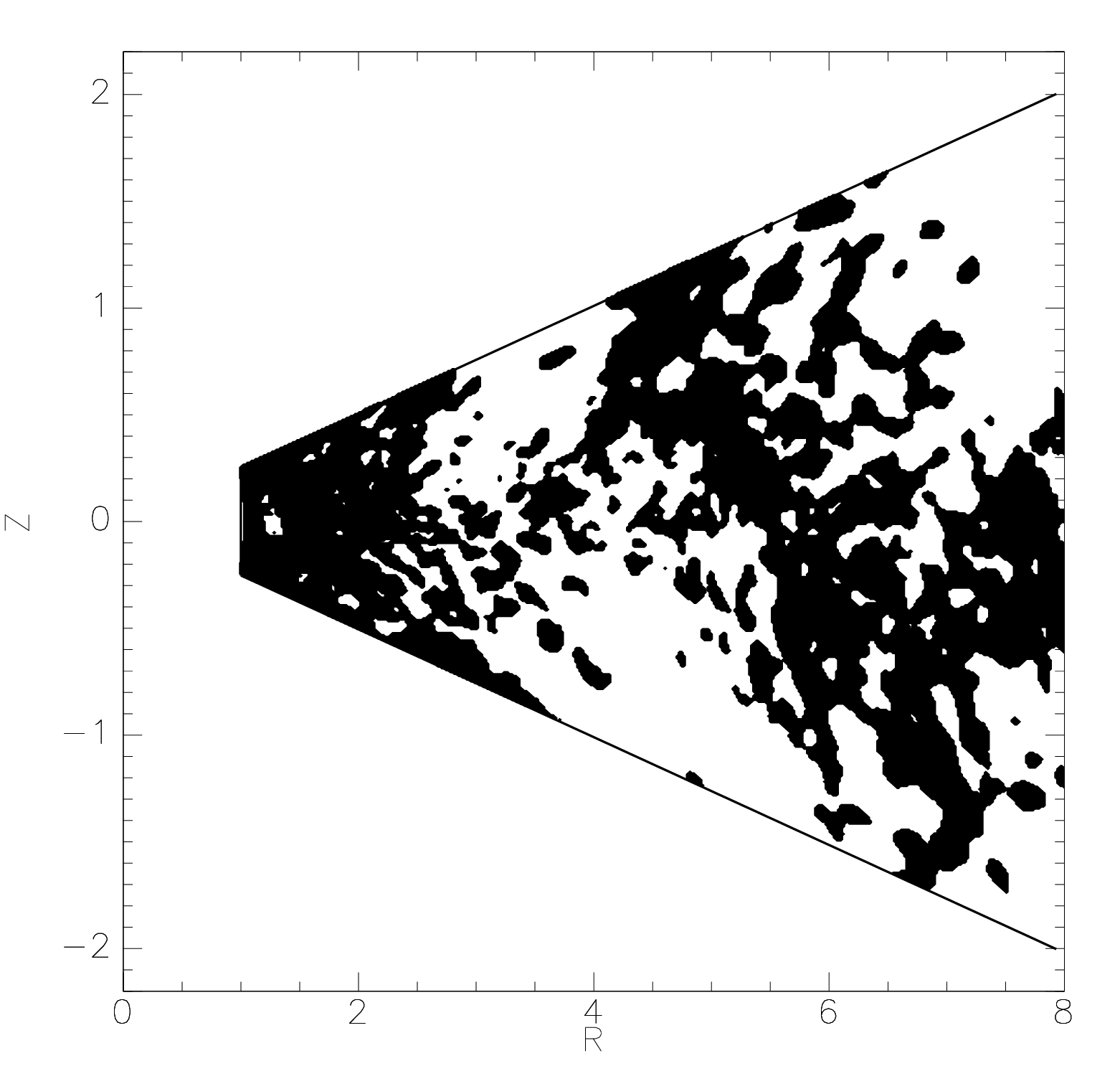}
\includegraphics[scale=0.45]{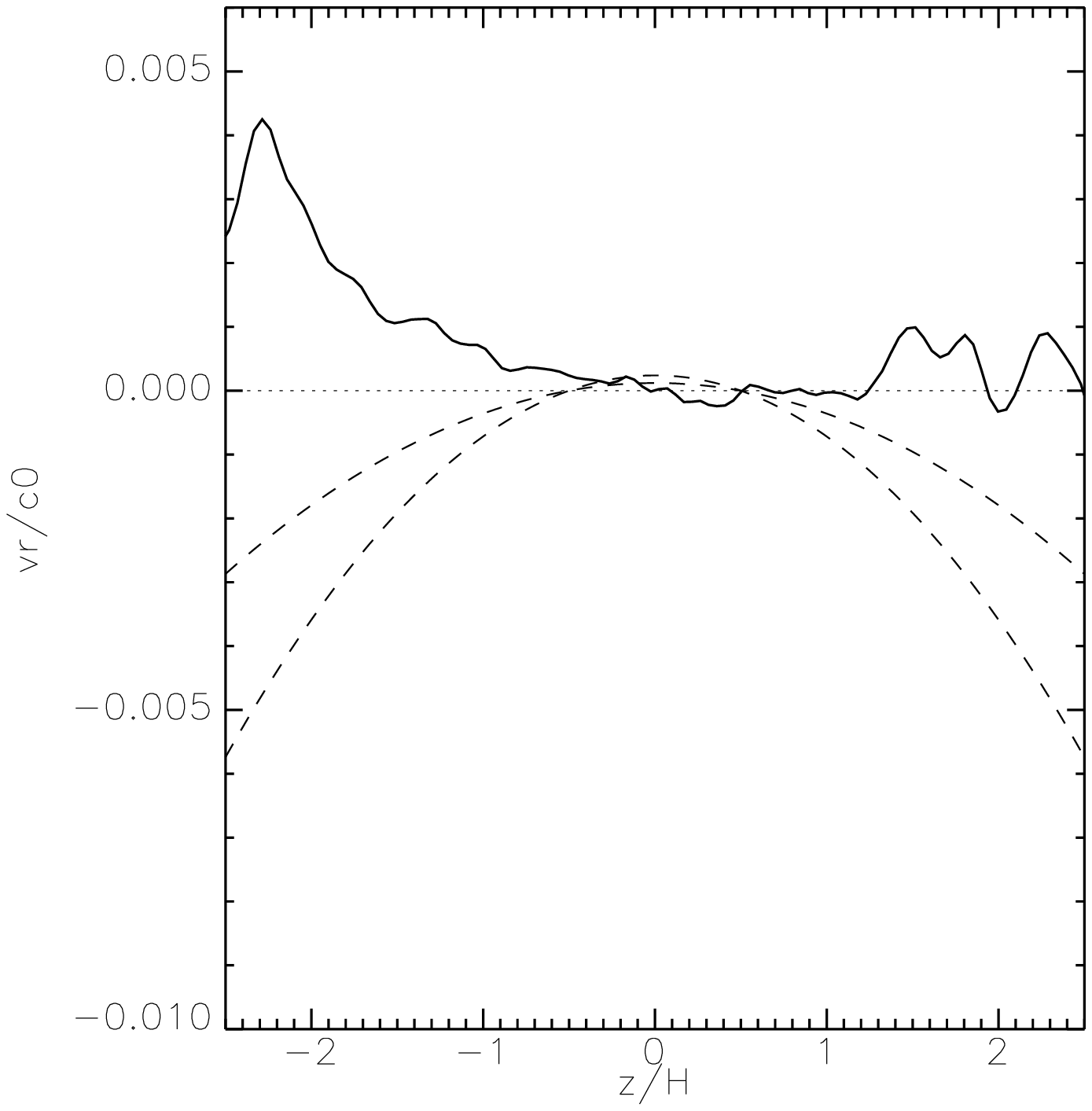}
\caption{Same as figure~\ref{case_p2_q1_fig} but for the case $p=-3/2$.}
\label{case_p1.5_q1_fig}
\end{center}
\end{figure*}

Even though the previous results suggest a significant difference
between viscous and turbulent models, the
simulations presented above still display large
fluctuations. To assess the robustness of the result, namely
the absence of any meridional circulation in turbulent disks, we
present in this section two additional simulations for which $p=-3/2$
and $p=-5/2$. These two runs serve two purposes: first, they
probe different disk radial structure. Second, and maybe more
importantly, they test the nature of the large--scale flow in disks
given different realizations of the turbulent flow.

The results of these two simulations are displayed in
figures~\ref{case_p2.5_q1_fig} and \ref{case_p1.5_q1_fig}
for the case $p=-5/2$ and $p=-3/2$, respectively. Both figures are
similar to figure~\ref{case_p2_q1_fig}: the left hand panel shows the
meridional 
distribution of $v_R$, with white regions corresponding to outward
motions and black regions to inward flow; the right hand panel plots the
vertical profile of $v_R$ radially averaged between $R=3$ and
$R=6$. This is compared to the predictions of viscous disk 
theory (depicted using a dashed line). Both models confirm the
differences between turbulent and viscous disks. No meridional
circulation is observed, and the results are similar to the
fiducial model $p=-2$. The absence of meridional circulation in
turbulent protoplanetary disks thus appears quite general
(i.e. independent of the disk structure).

\section{Numerical checks}
\label{num_checks_sec}

\begin{figure}
\begin{center}
\includegraphics[width=\columnwidth]{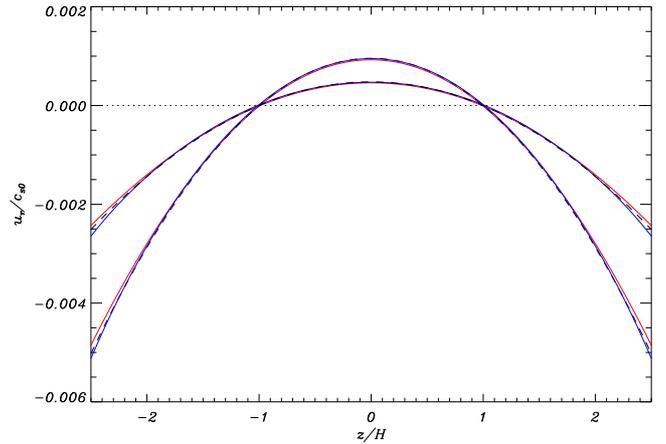}
\caption{Vertical profile of the gas radial velocity obtained in two
  hydrodynamical viscous simulations with the {\sc Pencil code} ({\it
    red solid line}) and with the JUPITER code ({\it solid blue
    line}). The two sets of curves plot the simulations results
  obtained with $\alpha=5 \times 10^{-3}$ and $10^{-2}$ while the
  dashed lines are the analytical predictions of
  Eq.~(\ref{vr_viscus_th}). Numerical and analytical results are in
  excellent agreement with each other.}
\label{num_checks_fig}
\end{center}
\end{figure}

The MHD simulations presented in the section above suggest that 
meridional circulation is not present in turbulent protoplanetary
disks. However, because of its very small amplitude,
one has to be careful since a number of potential numerical issues 
might affect this conclusion. For example, the grid meridional
boundaries might cause wave reflections that could perturb the
development of that circulation, while the limited radial extent of
the computational domain perturbs its viscous evolution. Both problems
can 
potentially affect the development of meridional circulation.
To address these issues, we have used two hydrodynamic codes,
namely the {\sc Pencil Code}\footnote{See 
  \url{http://www.nordita.org/software/pencil-code}} and the code
JUPITER \citep{devalborroetal06}. Both codes were used to solve the
hydrodynamic equations in spherical geometry with a prescribed
kinematic viscosity. The implementation of the viscous force in both
codes was checked using the test described in
Appendix~\ref{test_viscous}. At 
$t=0$, a disk in hydrostatic equilibrium is initialized as described in
section~\ref{disk_setup_sec}, using $q=-1$, $p=-2$, and
$H_0/R_0=0.1$. The $\alpha$ prescription is used with the two different values
$\alpha=5 \times 10^{-3}$ and $10^{-2}$. The size of the grid and the
resolution in the disk's meridional plane are identical to the MHD runs
performed with GLOBAL. However, the boundary conditions
differ. The {\sc Pencil} simulation was done with periodic boundaries 
in $\theta$. JUPITER in turn used simple reflexive boundary
conditions both in colatitude and radius, 
so that the ghost zones are filled with copies of the neighboring
zones of the active mesh, using a mirror symmetry.
This yields a trivial Riemann problem at the
boundary, hence a zero mass flux and a momentum flux that scales with
the pressure at the boundary.

The results obtained with the two codes are summarized in
figure~\ref{num_checks_fig}, in which the vertical profile of the
radial velocity is plotted for the two values of $\alpha$ quoted
above. All curves are averaged in azimuth and in radius between
$R=3$ and $R=6$. They were obtained at times $t=250$ ({\sc Pencil
  Code} results), $t=37.5$ (Jupiter results, case $\alpha=5 \times 
10^{-3}$) and $t=57.5$ (Jupiter results, case $\alpha=10^{-2}$). For
both codes and both values of 
$\alpha$, the agreement between the numerical results and the
analytical predictions is spectacular. There is only a tiny mismatch
above $2H$, which is most likely due to the meridional boundary conditions. This
remarkable agreement shows that meridional circulation, when it
exists, is a robust feature of the flow that is easily recovered in
numerical simulations. It 
is neither affected by the presence of artificial boundaries nor 
strongly influenced by the details of the numerical
algorithm. In addition, these results are obtained at much
shorter times than the disk's viscous timescale at $R=4.5$. Indeed the later
amounts to a few hundreds orbits, while the results are obtained after
a few tens of orbital times at that radius. This confirms the claims of
\citet{kley&lin92} and \citet{rozyczkaetal94} that meridional
circulation is established quickly, i.e. on a much shorter timescale 
than the disk's viscous timescales. As a conclusion, these viscous disk
numerical simulations give
further confidence that the lack of meridional circulation
reported in section~\ref{mhd_simus} is a real feature of turbulent
protoplanetary disks.

\section{Discussion}
\label{discussion_sec}

The numerical simulations presented above have shown striking 
differences between viscous and turbulent disks. In this section, we
first present a simple model that accounts for the numerical results
and discuss the relationship between these results and 1D standard
$\alpha$ disk models before highlighting the consequences of our findings
for large--scale radial transport of solids in protoplanetary disks.

\subsection{A simple model}
\label{turb_disk_model}

\begin{figure}
\begin{center}
\includegraphics[scale=0.5]{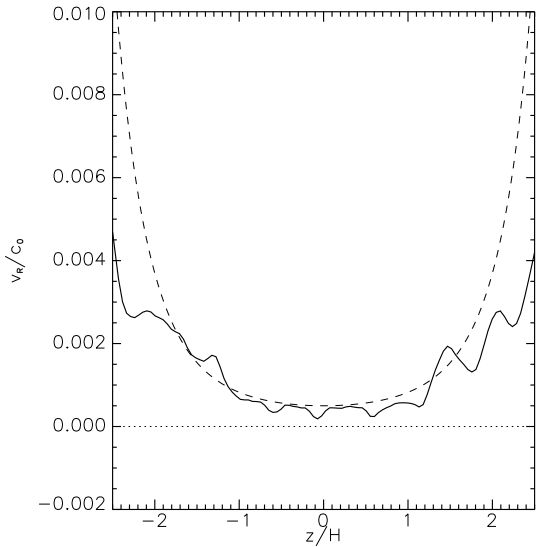}
\caption{The solid line shows the vertical profile of $v_R$ for the
case $p=-2$, derived as described in figure~\ref{case_p2_q1_fig}. It
should be compared with the dashed line which plots the prediction of
Eq.~(\ref{vr_turb_theory}), using $\alpha_t=5 \times 10^{-3}$
and $H_0/R_0=0.1$.}
\label{1d_vr_p2_th_fig}
\end{center}
\end{figure}

\begin{figure}
\begin{center}
\includegraphics[scale=0.5]{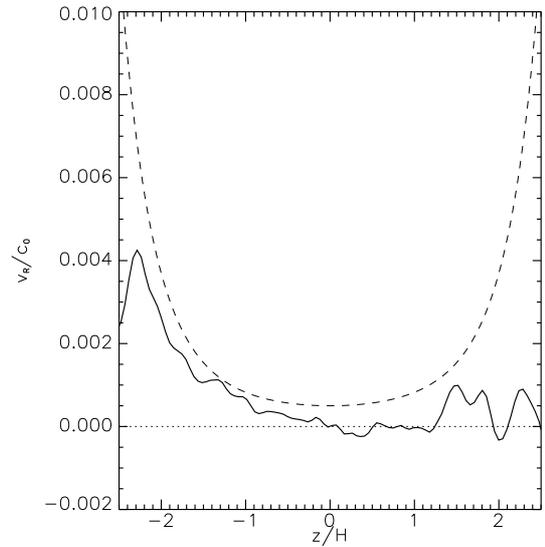}
\includegraphics[scale=0.5]{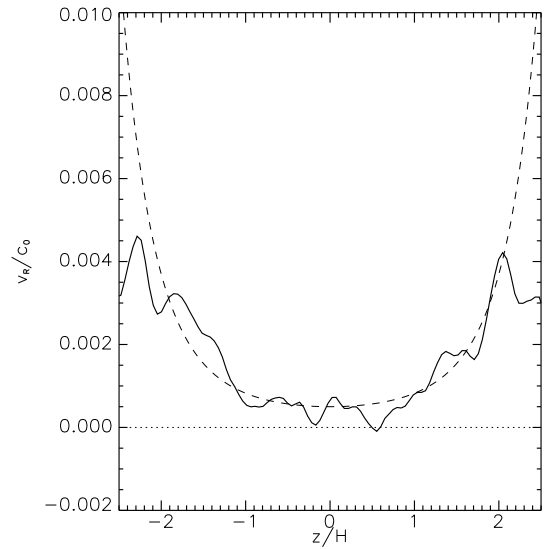}
\caption{Same as figure~\ref{1d_vr_p2_th_fig}, but for the cases
$p=-3/2$ ({\it top panel}) and $p=-5/2$ ({\it bottom panel}).}
\label{1d_vr_other_th_fig}
\end{center}
\end{figure}

In section~\ref{torque_sec}, it was shown that meridional circulation
fails to be established in turbulent protoplanetary disks. As explained in
section~\ref{torque_sec}, this behavior, unlike that of
viscous disks, is due to differences between the
viscous and turbulent stress tensors that are responsible for angular
momentum transport. In this section, we provide a simple prescription
for the turbulent stresses $T_{R\phi}^{turb}$ and $T_{Z\phi}^{turb}$
that accounts for the 2D structure of the large--scale
flow observed in  the simulations.

\noindent
First, the right hand panel of
figure~\ref{1d_stress_p2_fig} shows that the $R\phi$ component of the
turbulent and viscous stress tensors are comparable in amplitude. In
addition, both stresses are larger by about an
order of magnitude than the $Z\phi$ components of the stress
$T_{Z\phi}^{visc}$ and $T_{Z\phi}^{turb}$ (compare the vertical
scales on the two panels of that figure). This suggests that the
$Z\phi$ component of the stress is not a fundamental piece in the
process of setting the flow topology. For 
example, in the derivation of the radial velocity in
section~\ref{ang_mom_visc_sec}, the $R\phi$ contribution of the
viscous stress to the meridional variations of the radial velocity
writes as
\begin{eqnarray}
\lefteqn{\frac{v_R^{(R\phi)}}{c_0}=-\alpha \left( \frac{H_0}{R_0} \right) \left(
\frac{R}{R_0} \right)^{q+1/2}} \nonumber  \\
& & \hspace{2cm} \left[ 3p+3q+6 + \frac{3q+9}{2} \left(  \frac{Z}{H} \right)^2\right] \, .
\label{vr_viscus_th_nozphi}
\end{eqnarray}
This last expression is very similar to the complete expression for
$v_R$ that is given by Eq.~(\ref{vr_viscus_th}) and still indicates
outward motion in the disk midplane and inward motion in the disk
upper layers for a wide range of $p$ and $q$ values. Obviously, the
essence of the meridional circulation lies in the $R\phi$ component of
the viscous stress. We thus adopt, for simplicity, the following
prescription for the $Z\phi$ component of the turbulent stress:
\begin{equation}
T_{Z\phi}^{turb}=0 \, .
\label{tzphi_prescription}
\end{equation}
The conservation of angular momentum, Eq.~(\ref{ang_mom_turb_eq}), can
thus be written as
\begin{equation}
R\rho v_R\frac{\partial R^2 \Omega_K}{\partial R} =
\frac{\partial}{\partial R} \left( R^2 T_{R \phi}^{turb} \right) \, ,
\label{ang_mom_nozphi}
\end{equation}
where the vertical velocity has been dropped as it is an order $H/R$
lower than the radial velocity. Similarly, the vertical variation in
$\Omega$ appearing on the left hand side of the equation above has
been neglected because it would result in terms of order $(H/R)^2$, i.e. much
smaller than the vertical variations arising as a result of the
vertical density stratification. Using the variations for
the gas density given by Eq.~(\ref{rho_rz_gauss}),
Eq.~(\ref{ang_mom_nozphi}) can then be written as 
\begin{eqnarray}
\lefteqn{\frac{v_R}{c_0} = 2 \left( \frac{H_0}{R_0} \right) \left(
  \frac{R}{R_0} \right)^{-p-1/2} \frac{\partial}{\partial R}\left(
  R^2 T_{R \phi}^{turb} \right) } \nonumber
\\
&& \hspace{4cm} \frac{1}{R_0\rho_0 c_0^2} \exp \left(
  +\frac{Z^2}{2H^2} \right) \, . 
\label{vr_turb_theory_ini}
\end{eqnarray}
In the bulk on the disk ($|Z|<2.5H$), the results of the simulations
(figures~\ref{1d_stress_p2_fig} and
\ref{2d_stress_p2_fig}) demonstrate that $T_{R \phi}^{turb}$ is
largely independent of $Z$. Thus the only dependence with $Z$ in the
expression above is in the exponential, which explains why the sign of
the radial velocity does not depend on the distance to the midplane. 

Whether the radial velocity is positive or negative instead
depends on the radial derivative of the turbulent stress tensor. If the radial 
variations of $T_{R\phi}^{turb}$ are steeper than $R^{-2}$, $v_R$ is
negative everywhere. In the opposite situation, it is
positive, as found in the simulations presented in this paper. For
illustrative purposes, we adopt in the following a simple power--law radial
dependence for $T_{R\phi}^{turb}$ and writes:
\begin{equation}
T_{R\phi}^{turb}=\left\{ \begin{array}{ll} -\alpha_t \rho_0 c_0^2
    \left( \frac{R}{R_0} \right)^{\delta} & \textrm{if } |Z|<2.5H \\
0 & \textrm{otherwise} \end{array} \right. \, .
\label{trphi_prescription}
\end{equation}
Such a prescription bears similarities to the one adopted by
\citet{kretke&lin10} in their disk models. In Eq.~(\ref{trphi_prescription}),
$\alpha_t$ is a normalizing factor of the
stress that shall not be thought of as the standard $\alpha$
parameter. However, on dimensional grounds, $\alpha_t$ and $\alpha$
are expected to have the same order of magnitude. The power--law 
exponent $\delta$ is an unknown number that depends a priori on the disk
properties (for example the density and sound speed exponents $p$ and
$q$). It is important, at this stage, to stress that such a form in
the radial variations of the stress tensor has absolutely no
theoretical grounds and should only be viewed as a means to pushing
the analysis further. Using the expression above for the stress
tensor, the radial velocity of the gas can be written as 
\begin{eqnarray}
\lefteqn{\frac{v_R}{c_0} = - 2 \alpha_t (\delta+2) \left( \frac{H_0}{R_0} \right) \left(
  \frac{R}{R_0} \right)^{\delta-p+1/2}  \exp \left(\frac{Z^2}{2H^2}
\right) }
\label{vr_turb_theory}
\end{eqnarray}
when $|Z|<2.5H$, while $v_R$ vanishes in the disk's upper layers. The
results of the simulations presented in this paper show that $v_R$ 
is positive in the disk midplane, suggesting $\delta<-2$ for the set
of disk parameters probed in the present paper. 
It would be tempting at this stage to take the actual measured radial
variations in the stress instead of using the simple power law given
by the equation above. We chose not to do it for two main reasons:
first, as detailed in the following sections, these variations
may be polluted by numerical artifacts and thus may not be
representative of an actual protoplanetary disk, and second, whether a
meridional circulation develops does not depend on those 
variations, as shown by Eq.~(\ref{vr_turb_theory_ini}). 
Rather, we simply observe that the results of the
simulations indicate a midplane radial velocity on the order of $5 \times
10^{-4} c_0$. In other words, it is around $\alpha
(H_0/R_0)$. This suggests in turn that $2(\delta+2)
(R/R_0)^{\delta-p+1/2}$ is close to unity. Eq.~(\ref{vr_turb_theory})
for the radial velocity is compared with the simulation 
results, for the case $p=-2$, in the left hand panel of
figure~\ref{1d_vr_p2_th_fig}. In
computing $v_R$ using that equation, we have used $\alpha_t \sim \alpha
\sim 5 \times 10^{-3}$, $H_0/R_0=0.1$ while the remaining terms (with
the exception of the exponential!) are assumed to be equal to
one. Given the simplicity of the prescriptions detailed above, the
agreement between the simulation results and the prediction of
Eq.~(\ref{vr_turb_theory}) is very good.

Finally, the prediction of
Eq.~(\ref{vr_turb_theory}) is further compared with the simulation results
for the case $p=-3/2$ ({\it top panel}) and $p=-5/2$ ({\it bottom
  panel}) in figure~\ref{1d_vr_other_th_fig}. Again, the agreement
between the simulations and the prediction of
Eq.(\ref{vr_turb_theory}) is satisfactory for the case $p=-5/2$, while
the case $p=-3/2$ shows larger, but still modest deviations from the
simple analytical prediction. 

\subsection{Comparison with the predictions of standard $\alpha$ disk
  models}
\label{comp_1D_alpha_sec}


Standard 1D $\alpha$ disk theory considers the height and azimuthally
integrated version of Eq.~(\ref{ang_mom_visc}) to describe angular
momentum transport. (All $Z$--derivatives disappear in that
procedure.) Using the notation introduced in
section~\ref{flow_prop}, it is expressed as \citep{pap&lin95,balbus&pap99}   
\begin{equation}
R\Sigma \overline{v_R} \frac{d}{dR}\left( R^2 \Omega \right)= \frac{d}{dR} \left(
  R^2 \Sigma \overline{T_{R\phi}^{visc}/\rho} \right) \, .
\label{vert_avg_angmom}
\end{equation}
Assuming it has the form of a standard viscous stress tensor, the
height and azimuthally integrated $R\phi$ component of the stress are
then 
\begin{equation}
\Sigma \overline{T_{R\phi}^{visc}/\rho}=\frac{1}{2\pi}\int T_{R\phi}^{visc}
d\phi dz = \frac{1}{2\pi}\overline{\nu} \Sigma R \frac{d\Omega}{dR} \, . 
\label{vert_avg_trphi}
\end{equation}
Standard $\alpha$ disk models then make the ansatz that it is in fact
the vertically and azimuthally average kinematic viscosity
$\overline{\nu}$ that takes 
the form given by Eq.~(\ref{alpha_prescription}). This is different
from the theory developed in section~\ref{ang_mom_visc_sec}, where this
relation is assumed to hold locally. Using that prescription for
$\overline{\nu}$ in Eq.~(\ref{vert_avg_trphi}) and given the radial
dependence of the Keplerian angular velocity, the vertically and
azimuthally averaged stress tensor writes as
\begin{equation}
\Sigma \overline{T_{R\phi}^{visc}/\rho}=\frac{1}{2\pi}\int T_{R\phi}^{visc}
d\phi dz = -\frac{3}{2} \alpha \Sigma c_s^2  \, .
\end{equation}
For locally isothermal disks such as those considered in the present
paper, this relation can be equivalently expressed in terms of the
thermal pressure as
\begin{equation}
\int T_{R\phi}^{visc} d\phi dz = -\frac{3}{2} \alpha \int P d\phi dz \, .
\label{alpha_rel_1d}
\end{equation}
This equation shows that the relation between stress and pressure
holds for vertically integrated quantities in standard $\alpha$ disk
theory, while the model developed in section~\ref{large_scale_flow}
explicitly assumes that it holds locally:
$T_{R\phi}^{visc}=-3/2 \alpha P$. The results presented in this paper have
shown that such a local relation does not hold between the {\it
turbulent} stress and the thermal pressure, as it would otherwise
result in meridional circulation developing in turbulent
disks. Nevertheless, this does not mean that this relation would not
hold between the vertically averaged turbulent stress and the
vertically averaged thermal pressure, as is the case in standard
$\alpha$ disk theory as shown by Eq.~(\ref{alpha_rel_1d}). 

A first
attempt at answering this question can be made by comparing the
vertically and azimuthally averaged radial velocity predicted, on one
side, by 1D $\alpha$ disk model and, on the other hand, by the
numerical simulations of protoplanetary disks presented in this
paper. In standard 1D $\alpha$ disk models, the radial profile of the 
vertically integrated radial velocity can be computed by plugging
Eq.~(\ref{alpha_rel_1d}) into Eq.~(\ref{vert_avg_angmom}) to get
\begin{equation}
\frac{\overline{v_R}}{c_0}=-3 \alpha \left(
  \frac{H_0}{R_0} \right) \left( \frac{R}{R_0} \right)^{q+1/2} \left[
  p+\frac{3q}{2}+\frac{7}{2} \right] \, .
\label{vr_viscus1d_th} 
\end{equation}
Equation~(\ref{vr_viscus1d_th}) can also be recovered from
Eq.~(\ref{vr_viscus_th}) by a simple vertical
integration \citep{takeuchi&lin02}. For the value $q=-1$
considered in the present paper, it vanishes and changes sign for
$p=-2$. When $p<-2$, Eq.~(\ref{vr_viscus1d_th}) predicts a positive
vertically integrated radial velocity, while $\overline{v_R}$ is negative for
$p>-2$. This behavior is different from what is found in the
simulations, in which outward bulk motions were observed regardless of
the value of $p$. This suggests that turbulent disk behave
differently than standard 1D $\alpha$ disk models.


The systematic outward velocities obtained in the simulations can be traced
back to $\alpha$ not being a constant function of R, but
rather systematically decreasing outward. Indeed, when using the prescription
for the turbulent stress tensor given by Eq.~(\ref{trphi_prescription}),
$\alpha=\overline{T_{R\phi}^{turb}/\rho}/\overline{c_s^2}$ as given
by Eq.(\ref{alpha_eq}), is found to scale like
$R^{\delta+2}$. Constant $\alpha$ values thus correspond to disk
regions where $\delta=-2$. In constrast,
$\delta<-2$ in those parts of the disks where $\alpha$ decreases with
R. Using Eq.~(\ref{vr_turb_theory}), we can thus predict outward midplane
radial velocities in the region where $\alpha$ decreases radially (since
$\delta+2<0$ there). The
tendenct of $\alpha$ to decrease with $R$ in our simulations thus
explains the trend toward finding positive midplane radial velocities. As shown by
figure \ref{flow_prop_fig}, in model $p=-2$, 
$\alpha$ is almost constant in the region $R \in [4.5,7]$. In this
region, we thus expect to have $\delta \sim 2$, and
Eq.~(\ref{vr_turb_theory}) predicts a vanishing radial velocity in the
disk midplane. This is again consistent with the simulation results,
as shown in the right hand panel of figure
\ref{case_p2_q1_figII}. Finally, Eq.~(\ref{vr_turb_theory}) predicts
that negative midplane radial velocities are obtained when $\delta>-2$,
or equivalently when $\alpha$ increases outward. We note that this
prediction is consistent with the recent results of
\citet{flocketal11}. In simulations similar to those presented here,
these authors indeed reports radially increasing $\alpha$ and inward
radial velocities.

The systematic positive radial velocities we report here might
appear suspicious for the long term evolution of an accretion disk, so
it deserves a comment. We start by 
stressing that $\alpha$ disk models such as those used here also
exhibit that property for some 
combination of the exponents $p$ and $q$. This is the case for example
of the model $q=-1$ and $p=-2.5$, for which Eq.~(\ref{vr_viscus1d_th})
predicts a positive velocity implying outward mass transport. This can
only be a transient situation. The mass accretion rate 
$\dot{M}=-2 \pi R \Sigma v_R$ is negative in this case and scales like
$R^{p+3q/2+3}$. Thus its exponent is negative, showing that the mass
accretion rate and its derivative both decrease with $R$ in amplitude. On
viscous timescales, the radial profile of $\Sigma$ will flatten and
accretion will proceed. Ultimately, mass will fall onto the star as
angular momentum is transported outward. Although it is currently
impossible to run 3D MHD simulations on viscous disk timescales, the
same is most likely true of the turbulent disk that is simulated in
the present paper. We would thus expect to recover accretion if the
simulations were run long enough. However, this does not prevent a
comparison between viscous models and simulations when disks are not
in steady state, because the mean flow in disks is established on
a shorter timescale than the viscous timescale. The point of the present
section is precisely to highlight that the predictions of standard
$\alpha$ disk theory differ from the findings of numerical
simulations. That being said, we caution that these simulations only
probe such differences for the limited time and parameters enabled by 
current computational resources. Whether they are
general is beyond the scope of this paper and should be the focus of
future work.


We finally comment that the differences we report between turbulent
and standard $\alpha$ disk model are consistent with previous results
in the literature. For example, \citet{sanoetal04} report 
a scaling of the turbulent stress tensor as $P^{1/4}$, and not as $P$
as in standard $\alpha$ disk theory. This is consistent
with the simulations of \citet{lyraetal08}. Along the same lines,
\citet{hiroseetal09} have also recently reported the lack of a direct
relation between stress and thermal pressure using shearing box
simulations of turbulent disks. At the same time, the results presented in the
present paper regarding this question should be taken with
care. Recent studies have shown that the saturation level of the
MRI in the unstratified shearing box is strongly affected by
explicit dissipation
\citep{lesur&longaretti07,fromangetal07,simon&hawley09}. In the
simulations presented here, all the dissipation is of numerical 
origin. Thus numerical dissipation alone could mask any scaling of the
stress with the disk parameters. To make things even worse, numerical
dissipation in these simulations is in principle a function of
position since the effective radial resolution per disk
scaleheight varies with radius. Any definite conclusions regarding the
precise scaling of the vertically averaged stress with pressure should
thus wait for better control of dissipation in global simulations,
a difficult task by definition. The present results
highlight once more the need for an accurate determination of the
saturation of the turbulent stress as a function of the disk
parameters. Such a task is  beyond the scope of the present paper but
should also be the focus of future work.

\subsection{Conclusions}

Large--scale meridional circulation has been predicted in
protoplanetary disks based on 2D viscous disk theory. However, the flow
in a protoplanetary disk is known to be turbulent, most likely because
of the MRI. The question raised in this paper is thus simple:
does meridional circulation exist in turbulent protoplanetary disks?
This problem has been addressed using a set of numerical simulations of
turbulent protoplanetary disks. The large--scale radial flow of
the gas was computed by averaging the results in the azimuthal
direction, as well as on long time intervals. The results are found
to disagree with 2D viscous disk theory. There is no sign of any
meridional circulation in the disk. Instead, the sign of the radial
velocity is found to be constant in the bulk of the disk
($|Z|<2.5H$). This is shown to come from the vertical behavior of
the turbulent stress tensor. Instead of scaling like the local thermal
pressure as assumed in 2D viscous disk theory, the turbulent stress is
constant around the disk midplane before dropping in the disk corona,
where thermal and magnetic pressure become comparable. Such a vertical
profile of the turbulent stress has been found already in many
numerical simulations, regarless of their local and global nature. It
is thus a robust property 
of the flow in turbulent disks. In light of these properties of the
stress, the conclusion that no meridional circulation exists in
fully ionized and turbulent protoplanetary disks appears
unavoidable. 

This has important consequences for the transport of solids in the
solar nebula. The relevance of models in which crystalline dust radial
transport is largely based on meridional circulation can be questioned
\citep{keller&gail04,tscharnuter&gail07,ciesla07,ciesla09,hughes&armitage10}. In
addition, the simulations also raise the possibility of an outward
radial mass flux unexpected from a standard $\alpha$ disk model. This
opens up the possibility of an efficient mechanism to transport solid 
particles outward. Given the current state of the art of global
numerical simulations of protoplanetary disks, the viability of such a
feature of the flow nonetheless remains very uncertain. Future studies
of MRI--powered MHD turbulence 
should aim at better constraining the nature of the large--scale flow
in disks and the saturation amplitude of the MRI as a function of the
disk's properties (surface densities, temperature).

It is also important to realize that the simulations
presented here do not rule out the possibility of a large--scale
outward flow in protoplanetary disks. They only demonstrate it does
not exist in fully turbulent disks and that a viscous modeling of such
disks leads to erroneous results. There are a number of possibilities
by which a meridional flow might develop in real protoplanetary
disks. First, protoplanetary disks are known to be largely neutral and
most 
likely harbor a dead zone around their equatorial plane in which the
flow is stable to the MRI. While the exact size and dynamical status
of the dead zone is uncertain because it depends both on chemical
\citep{fromang02,ilgner&nelson06a,ilgner&nelson06b} and dynamical
effects \citep{fleming&stone03,turneretal07,ilgner&nelson08}, its very
existence in the planet--forming region seems difficult to avoid. The
large--scale flow in such disks has never been considered and might
lead to unexpected results that could affect the large--scale transport
of solid particles. For example, \citet{bai&stone10} recently showed
that the fast inward drift of large particles causes an outflow of gas
in the dead zone midplane that could carry small solids like CAIs
with it. Finally, one should bear in mind the idealized nature and the
limited sampling of the parameter space of the simulations presented
here. Additional physical ingredients (ambipolar diffusion, hall
effects) or different disk properties (temperature radial and vertical
profile, magnetic field configuration) might all change the vertical
profile of the stress tensor and create a large scale outflow in the
bulk of the disk.

\section*{ACKNOWLEDGMENTS}

The authors acknowledge insightful discussions with M.~Gounelle,
J.C.~Augereau, A.~Youdin, and G.~Lesur as well as a useful report from
an anonymous referee that helped clarify the non--steady state nature of
the disks in the simulations. The simulations presented in
this paper were granted access to the HPC resources of CCRT and CINES
under the allocation x2009042231 made by GENCI (Grand Equipement
National de Calcul Intensif).

\appendix
\section{Off-centered Keplerian viscous ring spreading in spherical geometry}
\label{test_viscous}
\begin{figure}
\begin{center}
\includegraphics[width=\columnwidth]{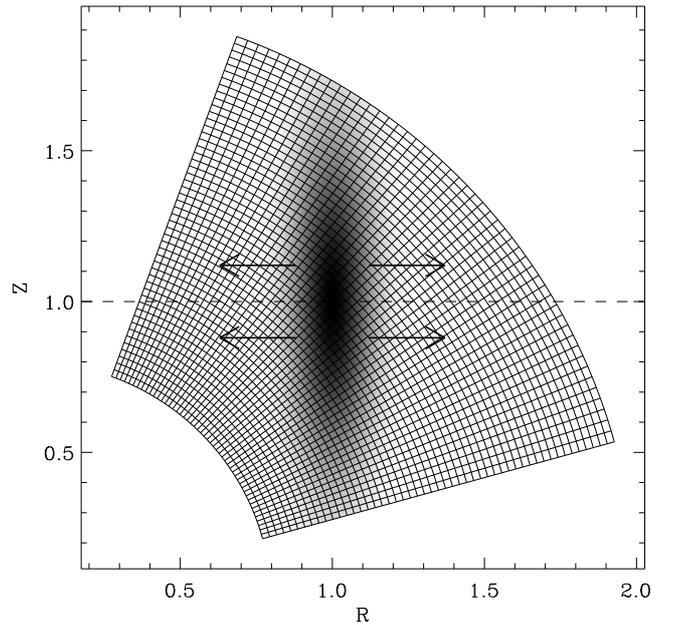}
\caption{Initial density field (gray levels) superimposed on the
  spherical mesh used in the calculation. The dashed line represents
  the ``equator'' (i.e. the potential minimum in the $Z$ direction),
  while the thick arrows depict the expected trend of the ring's
  material to spread radially.}
\label{fig:ap1}
\end{center}
\end{figure}

o
To test the implementation of the viscous stress tensor in
our purely hydrodynamical codes, we have designed a set up, in
spherical coordinates $(r,\theta,\phi)$, in which the two components
of the stress tensor that potentially contribute to meridional
circulation are within the same order of magnitude. It consists of a disk
orbiting within a fixed potential $\Phi$ that has the following form:
\begin{equation}
  \label{eq:1}
  \Phi(r,\theta,\phi) = -\frac{GM}{r\sin\theta}+\lambda(r\cos\theta-Z_0)^2,
\end{equation}
where $Z_0$ and $\lambda$ are constants.  This potential is Keplerian
along the cylindrical radius $R=r\sin\theta$, and harmonic in the
vertical direction $Z=r\cos\theta$, with a potential minimum at
$Z_0\ne 0$. An equilibrium setup for this potential can therefore
correspond, for a globally isothermal gas with sound speed $c_s$, to
an approximately
Keplerian disk with its equator lying at $Z=Z_0$ and with a uniform
thickness given by $H=c_s/\omega_z$, where
$\omega_z^2=\partial^2\Phi/\partial z^2=2\lambda$. Namely, we
initialize our disk as 
\begin{eqnarray}
  \label{eq:2}
  \rho(r,\theta,\phi) &=& \frac
  {m}{2\pi^2R_0^2H\sqrt{2\tau_0}}x^{-3/4}\times\\
&&\exp\left[-\frac{(x-1)^2}{\tau_0}-\frac{(Z-Z_0)^2}{2H^2}\right],\nonumber
\end{eqnarray}
where $x=R/R_0$ and $\tau_0=12\nu t/R_0^2$, which is the ratio of the
time and of the viscous timescale of the disk \citep{LP74}. $\tau_0$
is set to be much lower than unity in the expression
above. Equation~\eqref{eq:2} corresponds to a narrow Gaussian ring of
mass~$m$, centered on $R_0$, in vertical hydrostatic equilibrium, so
that it also has a Gaussian profile, centered on $Z_0$, in the vertical
direction. Eq.~\eqref{eq:2} stems from the classical expression of
\citet{LP74}, which describes the radial spread of an initially
infinitely narrow viscous ring, in which we approximate the Bessel
function $I_{1/4}(2x/\tau_0)$ with $\exp(2x/\tau_0)/\sqrt{4\pi
  x/\tau_0}$, using the fact that its argument $2x/\tau_0$ is large.
This means that we start with a ring that has already undergone some
viscous radial spread, although for a time short compared to its
viscous timescale, so that its width is small compared to its
radius. A meridional cut of the initial density field is represented
in Figure~\ref{fig:ap1}.

We initialize the azimuthal velocity as follows:
\begin{equation}
  \label{eq:3}
  v_\phi=\left\{\frac{GM_*}{R}-c_s^2\left[\frac 34+\frac{2(x-1)x}{\tau_0}\right]\right\}^{1/2},
\end{equation}
where the last two terms correspond to the rotational support provided
by the radial pressure gradient,
and we initialize the cylindrical radial velocity as
\begin{equation}
  \label{eq:4}
  v_R=\frac{\nu}{R}\left[\frac 34+\frac{6}{\tau_0}x(x-1)\right],
\end{equation}
from which we set $v_r=v_R\sin\theta$ and
$v_\theta=v_R\cos\theta$. Equation~\eqref{eq:4} stems from
Eq.~\eqref{ang_mom_visc_II}, in which the $Z$ derivative is set to
$0$, as no vertical dependence of $\Omega$ is expected in this
setup, each layer at a given altitude being a replica of the
equatorial layer, with a density ratio that depends on the altitude.

For our fiducial calculation, we adopt $c_s=10^{-3/2} (GM/R_0)^{1/2}$,
$\lambda=10^{-2}/3 (GM/R_0^3)$, hence $H=\sqrt{0.15}R_0$, and we start
with a narrow ring corresponding to $\tau_0=0.018$, with an equator at
$Z_0=1$. Our mesh contains $50\times 50$ zones, regularly spaced in
radius from $R=0.8R_0$ to $R=2R_0$, and regularly spaced in colatitude
from $\theta=0.35$ to $\theta=1.3$. Since the azimuthal velocities are
taken into account, our setup is ``2D1/2'' in nature.

\begin{figure}
\begin{center}
\includegraphics[width=\columnwidth]{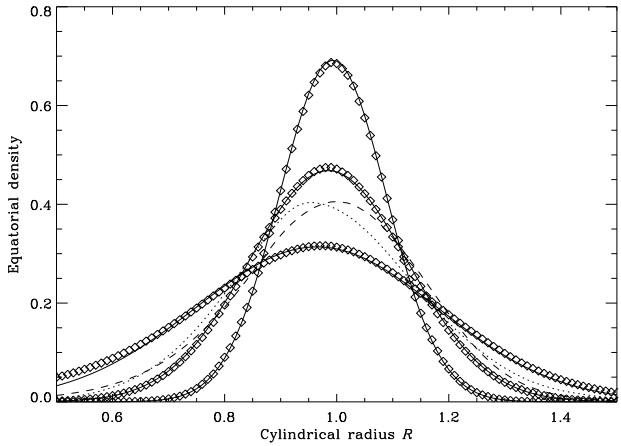}
\caption{Equatorial density cuts for $\tau_0=0.018$ (initial
  conditions, highest and narrowest peak), $\tau_0=0.04$ and
  $\tau_0=0.09$ (widest distribution). In each case the diamond's curve
  represents the distribution inferred in the $Z=Z_0$ plane from the
  numerical simulations, using a bilinear interpolation, while the
  solid curve represents the exact solution of \citet{LP74}. In
  addition, the dotted and dashed curves represent the situation
  obtained for $\tau_0=0.09$ in numerical calculations in which we set
  $T^{\rm visc}_{\theta\phi}$ and $T^{\rm visc}_{r \phi}$ to zero,
  respectively.  }
\label{fig:ap2}
\end{center}
\end{figure}
We present in Figure~\ref{fig:ap2} the results of this calculation, 
which show that the radial spread of the ring is correctly captured on
the spherical mesh. We performed subsidiary calculations in which
we manually cancel out one of the two stress components that have an
impact on the torque felt by a ring of material (hence on the rate at
which the material spreads), i.e. either $T^{\rm visc}_{\theta\phi}$
or $T^{\rm visc}_{r \phi}$. That the resulting curves of
these two additional calculations display a significantly different
result than the correct result and that they yield a similar peak
value at the same date shows that these two components of the stress
tensor play roles of comparable magnitude on the spread of the
ring. This can be understood as the spherical coordinate system, in
the vicinity of the center of the ring at $Z=Z_0$, is approximately
tilted by $45^\circ$ with respect to the equator.  We finally note
that the original derivation of \citet{LP74} does not contemplate
pressure effects. Those are relatively substantial in the disk
considered here, and might account for the tiny discrepancy between
the simulated and expected profiles in Figure~\ref{fig:ap2}, except
near the boundaries, where the discrepancy likely arises from the
accumulation of material due to the reflexive boundary conditions that
we use, which prevent material from flowing out of the grid.

\bibliographystyle{aa}
\bibliography{author}

\end{document}